\documentclass[fleqn,usenatbib]{mnras}

\usepackage{newtxtext,newtxmath, float, mathtools}
\usepackage{cuted}
\usepackage{xcolor}
\usepackage{orcidlink}
\usepackage[T1]{fontenc}

\DeclareRobustCommand{\VAN}[3]{#2}
\let\VANthebibliography\thebibliography
\def\thebibliography{\DeclareRobustCommand{\VAN}[3]{##3}\VANthebibliography}

\usepackage{graphicx}	
\usepackage{amsmath}	

\title[GLANCE -- Lensed GW Signal Finder]{GLANCE -- Gravitational Lensing Authenticator using Non-Modelled Cross-Correlation Exploration of Gravitational Wave Signals}
\author[Aniruddha Chakraborty, Suvodip Mukherjee]
{
Aniruddha Chakraborty\orcidlink{0009-0004-4937-4633}\thanks{E-mail: aniruddha.chakraborty@tifr.res.in },
Suvodip Mukherjee\orcidlink{0000-0002-3373-5236} \thanks{E-mail: suvodip@tifr.res.in }
\\
Department of Astronomy and Astrophysics, Tata Institute of Fundamental Research Mumbai, Mumbai-400005, Maharashtra, India
}

\date{Accepted XXX. Received YYY; in original form ZZZ}
\pubyear{2024}
\begin{document}
\label{firstpage}
\pagerange{\pageref{firstpage}--\pageref{lastpage}}
\maketitle

\begin{abstract}
Gravitational lensing is the phenomenon where the presence of matter (called a lens) bends the path of light-like trajectories travelling nearby. Similar to the geometric optics limit of electromagnetic waves, gravitational lensing of gravitational waves (GWs) can occur in geometric optics condition when GW wavelength is much smaller than the Schwarzschild radius of the lens i.e. $\lambda_{GW} \ll$ R$^{\rm s}_{\rm lens}$. This is known as the strong-lensing regime for which a multiple-image system with different magnifications and phase-shifts is formed. We developed \texttt{GLANCE}, Gravitational Lensing Authenticator using Non-modelled Cross-correlation Exploration, a novel technique to detect strongly lensed GW signals. We demonstrate that cross-correlation between two noisy reconstruction of polarized GW signals shows a non-zero value when the signals are lensed counterparts. The relative strength between the signal cross-correlation and noise cross-correlation can quantify the significance of the event(s) being lensed. Since lensing biases the inference of source parameters, primarily the luminosity distance, a joint parameter estimation of the source and lens-induced parameters is incorporated using a Bayesian framework. We applied \texttt{GLANCE} to synthetic strong lensing data and showed that it can detect lensed GW signals and correctly constrain the injected source and lens parameters, even when one of the signals is below match-filtered  threshold signal-to-noise ratio. This demonstrates \texttt{GLANCE}’s capability as a robust detection technique for strongly lensed GW signals and can distinguish between lensed and unlensed events.

\end{abstract}




\section{Introduction}

Gravitational waves (GWs) are the newest addition to the field of multi-messenger astrophysics. Since the first observation of the binary black-hole (BBH) merger on 14th September, 2015 \citep{PhysRevLett.116.061102}, there have been about 90 observations till their third observational run \citep{KAGRA:2023pio}. These ripples in the fabric of spacetime provide us with a new way to understand the universe: 
inferring the Hubble constant \citep{LIGOScientific:2021aug} , population studies of merging compact binaries \citep{PhysRevX.13.011048} , or testing the general theory of relativity  \citep{LIGOScientific:2021sio}. 
Aside from its many differences with EM waves, GW also propagates in a light-like trajectory and their path gets altered by intervening matter according to the general theory of relativity. This phenomenon is known as the gravitational lensing of GW \citep{PhysRevLett.77.2875, PhysRevLett.80.1138, Bartelmann_2010,  PhysRevD.95.044011}. Owing to its kilometer-to-petameter range of wavelengths, GW provides us with new scales to observe the large-scale universe: from dwarf galaxies to galaxy clusters in different lensing regimes. 

\begin{figure*}
   \centering
    \includegraphics[width=0.98\textwidth]{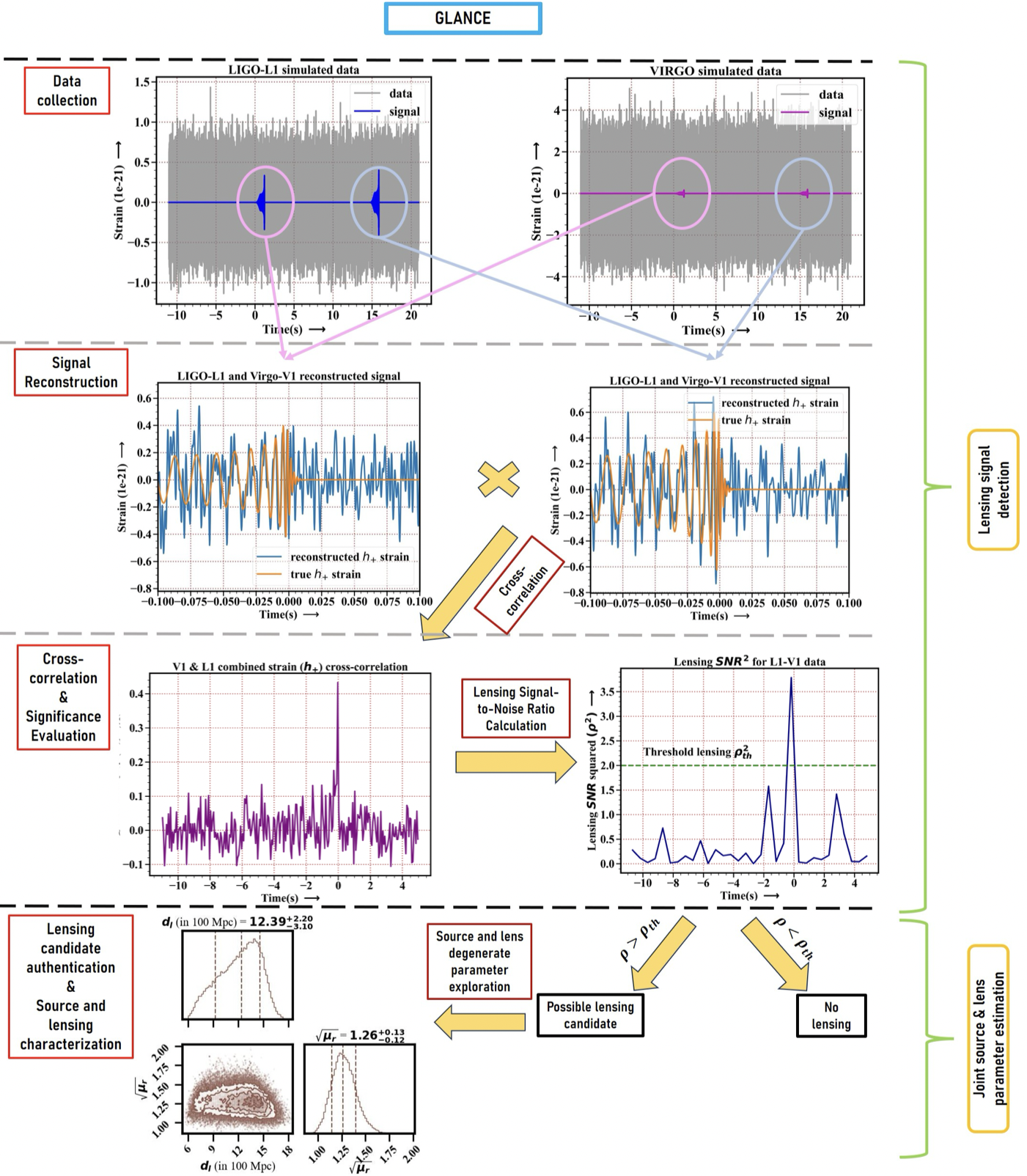}
    \caption{The figure shows the outline for the strong lensing searches of GW using \texttt{GLANCE}. We take the data from the GW observatories and extract the information about individual GW polarization as best as we can. We cross-correlate two such signals, separated by lensing time delay, containing only one polarization. The strength of the cross-correlation is measured by calculating the deviation signal cross-correlation from typical noise cross-correlation fluctuations. We call a pair of events as candidates for significant deviation of the signal cross-correlation from the noise cross-correlation. For those candidate events, we remove the lensing bias affecting the inference of the source parameters by estimating the parameters of the source jointly with the lens-induced parameters.} 
    \label{fig1}
\end{figure*}

Gravitational lensing of GW follows directly from Fermat's principle of least time in presence of intervening matter. Lensing can affect the spatial trajectory of GW leading to phase differences among GWs propagating along different paths. During interference between such phase-shifted waves, this creates an overall amplification of GW. Much like ray optics and wave optics for eletromagnetic waves there are two distinct areas of lensing of GW: (i) Geometric optics: in this limit, the Schwarzschild radius $(R^{\rm s}_{\rm lens})$ of the intervening lensing object, given by $R^{\rm s}_{\rm lens}=\frac{2GM_{\rm lens}}{c^2}$ is much larger than the wavelength $(\lambda)$ of the GW. Such cases are referred to as strong lensing. (ii) Wave optics: in this limit, the Schwarzschild radius of the lens $(R^{\rm s}_{\rm lens})$ is of comparable size or smaller than the wavelength $(\lambda)$ of the GW. This is known as the microlensing case. 
Due to the very different length scales involved with the wavelengths on GWs (from $10^5$m (LIGO observable) to $10^{17}$m (PTA observable)), the impact of gravitational lensing on GWs can be studied in both geometric-optics and wave-optics regimes (or in the transition between them) depending on the lens mass and radius. Multiband GW signal from compact objects covering nearly nano-Hertz to kilo-Hertz signals can have different observable signatures in geometric optics and wave optics limit making it possible to explore cosmic structures of masses ranging from $10^{14}$ M$_\odot$ to $10^3$  M$_\odot$ \citep{WAMBSGANSS2006567, Bayer_2023} approximately. In the absence of any luminous baryonic matter in the cosmic structures, lensing of GWs becomes the key method to study the electromagnetically undetectable dark matter structures over this mass range. The detection of lensed GW signals can lead to a plethora of unexplored science cases ranging from astrophysics to fundamental physics \citep{Congedo:2018wfn,Mukherjee:2019wfw, Mukherjee:2019wcg, Goyal:2020bkm, Basak:2021ten, Ezquiaga_2021, Mpetha:2022xqo,Balaudo:2022znx,Caliskan:2023zqm, Narola:2023viz}. 

The objective of this project can be divided into three stages, as follows: Firstly, the detection of lensed GW signal, which is the main dealing of this paper. Then we infer the source and lens-induced characterizations jointly and exploring their degeneracies. In the end, we can study the properties of the massive lensing objects from the data-driven approach given a handful of lensing observations.

In the current LIGO-VIRGO-KAGRA(LVK) sensitivity range \citep{Martynov_2016, KAGRA:2013rdx, Buikema_2020}, observing a lensed GW is a very rare event, it has the probability of occurrence of a few parts in a thousand (0.1 - 0.6\%) \citep{10.1093/mnras/sty411, 10.1093/mnras/stab1980, Diego_2021}. The lensing search by LVK on the O3 data led to no significant evidence of a lensed GW event \citep{2023arXiv230408393T} using existing methods \citep{10.1093/mnras/stab1991, Wright:2021cbn, PhysRevD.107.123015}. A follow-up study on the O3 events did not bring up any significant evidence of the observation of a lensing event so far \citep{10.1093/mnras/stad2909}. With the improved detector network in the upcoming O5 run, it is expected to observe at least one pair of strongly lensed events.

With the lensing observation chances increasing with improving detectors \citep{Abbott_2020}, the necessity of a robust lensed GW finder becomes significantly important, a technique which can detect these rare lensed events confidently and isolate them from non-lensed event with similar signal characteristics. 

Keeping this in mind, in this work, we have proposed a new and robust technique for the detection of lensed GW signals, \texttt{GLANCE: Gravitational Lensing Authenticator using Non-modelled Cross-correlation Exploration}. In fig. \ref{fig1} we depict the outline of the \texttt{GLANCE} framework. In a nutshell, we will use gravitational wave time-series data, chop it and extract the polarization information from it at a certain time and do the same for some other time. These two polarization information is then cross-correlated. If the cross-correlation is performed just at the right phase for both of those signals, we observe a peak in the cross-correlation signal. The strength of the peak cross-correlation is measured by calculating its deviation from typical noise cross-correlation. We call an event as a lensing candidate if for those pair of events the lensing signal-to-noise ratio (lensing SNR) is above some threshold. Since lensing biases the source properties by modifying those with some lens-imposed signatures on the signals. To infer their source parameters and to remove any false biases, we use both data pieces to characterize the source and lens jointly using Bayesian framework. Application of cross-correlation technique on GW data analysis is very useful to search for beyond modelled non-GR signatures imprinted on GW \citep{Dideron_2023}, stochastic GW background \citep{allen1999detecting, Romano_2017} and continuous GW searches \citep{Dhurandhar:2007vb}. 

The following sections are organized as mentioned here: I) \textbf{Salient aspects of \texttt{GLANCE}}: In this section, we are going to discuss the current scenario in lensing analyses and why a new pipeline is required to detect lensed GW signals. II) \textbf{Basics of GW generation from compact binary coalescence}: In this section, we will find out how gravitational waves are generated from compact binary mergers and how they propagate in free space. III) \textbf{Gravitational lensing identifier technique using cross-correlation}: In this section, we formally describe the mathematical framework of our proposed technique to detect a lensed signal for strong lensing case. IV) \textbf{Lens and source characterization of the strong lensing candidates}: In this section, we estimate the source properties and lens characteristics imprinted on the signal for a strong lensing candidate event.

\section{Salient aspects of \texttt{GLANCE}}
With increasing volume in the strain data from the existing GW observatories and the chances of the observation of lensing increasing with next-generation GW detectors, the necessity of having good tools to analyze it and search for lensing signatures feels very important. In this prospect, we have developed a new method of detecting and characterizing the lensed GW signals called \texttt{GLANCE}.  
Since GW is a highly coherent wave and any alternation in phase, if there, is an observable in the data channels, the phase information is a very clean probe to study lensing. Thus by applying cross-correlation between data pieces at the right phase of the GWs, we can find the degree of similarity between two signals as a novel technique to detect lensed GWs. From this perspective, \texttt{GLANCE} stands out from any current lensing finder technique.

\begin{itemize}
    \item \texttt{GLANCE} works across detectors. It can combine data from different detectors at different times to find any traces of lensing. The application of cross-correlation on the reconstructed polarization data then helps us to find lensed signals. For non-observation of a lensed signal by one of the detectors, we can use data from other detectors to make that event counted as a candidate for lensing.
    \item \texttt{GLANCE} performs cross-correlation on two noisy reconstructed one-polarization signals. It calculates the deviation of the time-averaged signal cross-correlation from typical time-averaged noise cross-correlation, weighted by the standard deviation of the noise correlation. This is known as lensing SNR. We qualify an event as a lensing candidate when the lensing SNR (or rather, the square of the lensing SNR) is above a certain cut-off value. This helps to quantify the similarity between two GW signals and thus helps two different signals to be authenticated as lensed counterparts.
    \item To remove lensing biases, \texttt{GLANCE} explores the joint parameter space comprising of the source properties and the lens-induced characteristics. To perform the joint parameter estimation, \texttt{GLANCE} incorporates a Bayesian framework. 
    
\end{itemize}

\subsection{Comparison with existing techniques}

In the third observational run (O3), LIGO-VIRGO-KAGRA (LVK) strong lensing search techniques primarily used the posterior overlap technique for lensing searches \citep{2023arXiv230408393T}. It is based on a bayesian inference to obtain the probability distributions of the source parameters (posteriors) for two potential lensing candidate events. It then calculates the overlap between those parameters unaffected by lensing (such as masses, spins, inclination angle). When the signal is loud, the posteriors of the event can be well constrained. However, if the signal is weak, the posteriors can be broad. So, if we are comparing two signals out of which one signal is a super-threshold and the other one is a sub-threshold, we can get overlap between posteriors even if they are not lensed counterparts of one another. So, posterior overlap technique cannot perform in the sub-threshold regime to look for strongly lensed GW signals. However, cross-correlation techniques like \texttt{GLANCE} can work even when one of the signals are in the sub-threshold regime.

Another inference technique used was the joint source parameter estimation (PE) \citep{janquart2022golum}. The method follows up from the posterior overlap for events showing significant posterior overlapping (of the parameters unaffected by lensing) and tries to infer the lens-imposed characteristics (such as relative magnification, Morse-phase) as well as the intrinsic source parameters (such as masses and spins) from the signal jointly from two lensing candidate data pieces. This joint-parameter estimation technique, although very reliably infers the relative magnification and phase-shift, suffers from a major drawback when the signals are weak and no tight constraints on the parameters are obtained.

Cross-correlation based techniques like \texttt{GLANCE}, observes strain-level overlap as compared to parameter overlap technique, in which the strain is first used to estimate relevant parameters and then calculation on the overlap between posteriors is performed. To compare, \texttt{GLANCE} performs a first order calculation by cross-correlating the data whereas, posterior overlap and joint PE techniques use a second order calculation on the data in obtaining the parameters first and then calculating the posterior overlaps. Therefore, sub-threshold or low match-filtering SNR strong lensing search techniques become ineffective using these techniques. `

Sub-threshold lensing search for finding strong lensed signals have also been performed \citep{Li2019TargetedSS}. The technique calculates a best-fit template for a super-threshold event (more magnified) and uses that template to search for sub-threshold lensed counterparts (less magnified). The template-based search assumes that the component masses and spins are unaffected by lensing-modifications on the waveform in the strong lensing limit. A shortcoming of the technique lies in its model-dependent. Any unmodelled signature present on the signal (whether due to lensing or due to some beyond general relativity signatures), may result in a search failure. On the contrary, \texttt{GLANCE} performs a model-independent search through the strain data to look for two similar signals without any assumption on the characteristics imposed by lensing. This makes our cross-correlation based strong lensing search a model-free search approach.

Recent advancements on the field use machine learning based techniques to search for lensed GW signals \citep{PhysRevD.104.124057}. These techniques try to observe the frequency-evolution (q-transformed signal) of the signals with overlapping sky-localizations, calculated using BAYESTAR \citep{Singer_2016}. The machine is trained to find the similarity between the noisy spectrograms. The machine-learning based technique is a model-based approach, so far spinning, precessing BBHs or eccentricity of the orbit has not been incorporated into that analysis. Also training the machine to perform lensing searches technique works for high-SNR events, but in the sub-threshold regime it becomes quite ineffective. In this aspect, \texttt{GLANCE} works in a model-independent approach and also it can accommodate any super-threshold and sub-threshold signals. Again, any non-Gaussianity in the noise (e.g. glitch) poses challenges for a technique purely looking into the similarity of the spectrograms. However, such noise artefacts are often uncorrelated across detectors, thus cross-correlation techniques are immune to glitches or any form of noise non-Gaussianity present in the data. \texttt{GLANCE} takes around two minutes for the polarization signals reconstruction (for data sampled at 4096Hz), their cross-correlation and its significance analysis in terms of the lensing SNR, whereas the machine learning techniques are faster, taking a few seconds for the analysis after the machine being trained already. However, at slightly longer timescales, the lensing false alarm rate (FAR) is lower in \texttt{GLANCE} in comparison to machine learning techniques (see section \ref{section6} for details), giving cross-correlation techniques an edge above the machine-learning based ones. As we will later see, the FAR of this technique is very small making it a very reliable technique to identify lensing.

This technique, thus, can play a pivotal role in the search for lensed GW signals and can add to the ongoing efforts in searching for a lensed event from the available GW data.

\section{Basics of Gravitational Wave Generation from coalescing compact binary objects}
Gravitational waves are the emissions of gravitational energy from the cataclysmic and violent events in the universe. It is the propagation of the spacetime metric perturbation in the form of a transverse wave. The generation and propagation of gravitational waves follow the Einstein field equations \citep{Einstein:1915ca}. Gravitational wave has two polarizations: plus ($h_{+}$) and cross polarization ($h_{\times}$). Total emitted gravitational wave is a contribution of these two linearly independent polarizations, $h = h_{+} + h_{\times}$

The generation of gravitational waves follows a very similar equation to the Poisson equation and thus its solution is as follows:
\begin{equation}
    h_{\mu \nu}(t, \vec{R})=\frac{4G}{c^4} \int \frac{T_{\mu \nu}(t-s/c, \vec{r}) d V}{s},  \text { where } s \equiv|\vec{R}-\vec{r}| \quad,
\end{equation}
where $T_{\mu \nu}$ is the energy-momentum stress tensor with $G$ is the gravitational constant and $c$ is the speed of light in vacuum. $\vec{r}$ is the position of the source and $\vec{R}$ is the position of the observer, with respect to a certain origin, therefore $\vec{s}$ is the position of the observer with respect to the source.
Using the approximation that the source is placed very close to the origin as compared to the observer i.e. $|\vec{R}| \gg |\vec{r}|$ and using the conservation law of stress-energy tensor, we can show that the spatial components of the spacetime metric perturbation tensor follow:
\begin{equation}
    h_{i j }(t, \vec{R}) = \frac{2G}{c^4 s} \ddot{I}_{i j} \left(t- \frac{s}{c} \right) \quad,
\end{equation} 
where, $I_{i j} \equiv \int \rho x_i x_j d V$ is the moment of inertia of the source ($x_i$'s be the source coordinates), and a dot above indicates its time derivative, hence $I_{i j}$ here is twice differentiated in time. 
\begin{figure*}
    \centering
    \includegraphics[width=0.9\textwidth]{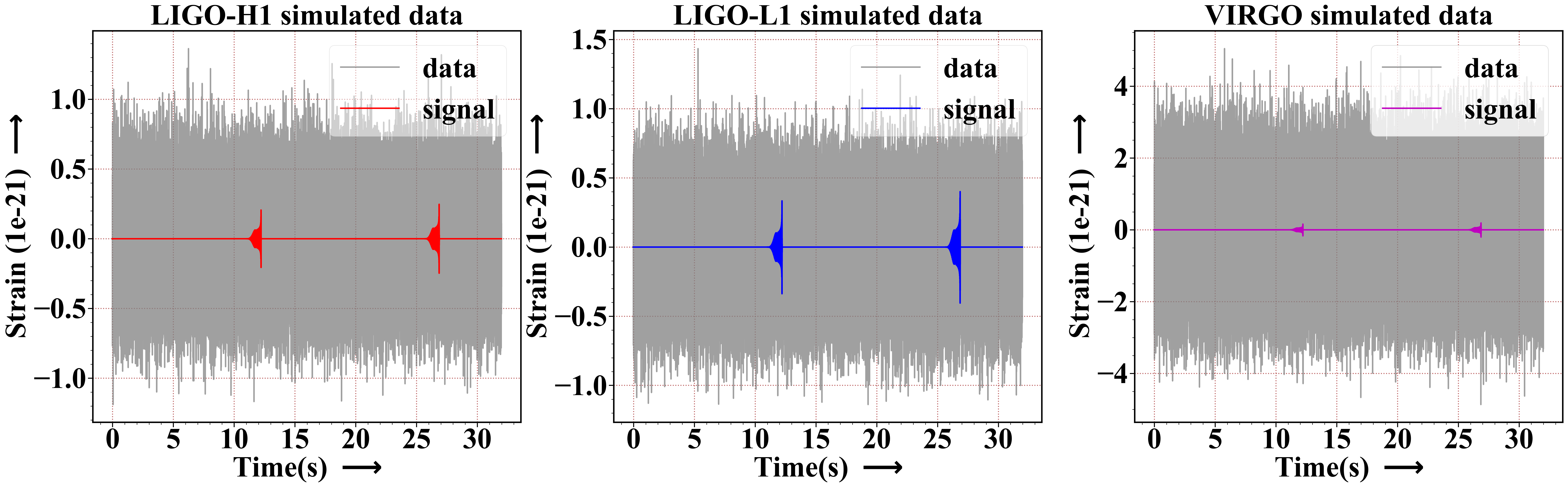}
    \caption{The figure shows the strongly lensed GW signals hidden behind the noise in LIGO-Livingston, LIGO-Hanford, VIRGO gravitational wave observatories.}
    \label{fig5}
\end{figure*}

Now let us consider a simple system to understand the GW emission from a coalescing binary black hole (BBH). We consider a BBH system with two black holes `a' \& `b' with both masses equal to $m$ revolving around their common center of mass in radius $R_0$. With origin chosen at their center of mass, the instantaneous coordinates of black hole (BH) `a' in the xy plane, 
$x_a = R_0 cos\Omega t $ and $y_a = R_0 sin\Omega t $ and the coordinates of BH `b', $x_b = -R_0 cos\Omega t $ and $y_b = -R_0 sin\Omega t $ where $\Omega$ be the angular velocity of each BH around the center of mass. We can find the time derivatives of the moment of inertia of this binary system as a function of time from the density \footnote{Approximation of black holes as point particles is very crude, but it develops a reasonable first-hand understanding.},
\begin{equation}
\begin{aligned}
\rho(t,\vec{x})= m\delta(z)[\delta(x-R_0 cos\Omega t)\delta(y-R_0 sin\Omega t) \\ 
+ \delta(x+R_0 cos\Omega 
t)\delta(y+R_0 sin\Omega t)].
\end{aligned}
\end{equation} 

We can thus obtain the gravitational wave strain as a function of space and time as, 

\begin{equation}
    h_{ij} (t, \vec{R}) = \frac{8Gm}{s} \Omega^2 R_0^2 \left[\begin{array}{ccc} -cos2\Omega t_{r} & -sin2\Omega t_{r} & 0 \\ -sin2\Omega t_{r} & cos2\Omega t_{r} & 0 \\ 0 & 0 & 0 \end{array}\right] \quad, 
\end{equation}
where, $t_r =t - s/c$ is the retarded time.

The frequency domain GW strain for a compact objects' merger with unequal masses, is a combination of many different modes with the dominant mode as $(l,m)=(2,2)$ (where l and m are the indices of the spherical harmonics $Y_{lm} (\theta, \phi)$), which is given by \citep{PhysRevD.52.848, PhysRevD.49.2658}, 
\begin{eqnarray}
    h_{+, \times}(f)(\hat{n})= \sqrt{\frac{5}{96}} \frac{G^{5 / 6} \mathcal{M}_z^2\left(f_z \mathcal{M}_z\right)^{-7 / 6}}{c^{3 / 2} \pi^{2 / 3} d_l} \mathcal{I}_{+, \times}(\hat{L} . \hat{n}),
\end{eqnarray}
where, we have accounted for the expansion of the universe with the source at a redshift of $z$, $f_z=f(1+z)$ is the redshifted frequency, $\mathcal{M}_{z}=\mathcal{M}(1+z)$ is the redshifted chirp mass (written in terms of the source frame chirp mass $\mathcal{M}=(m_1m_2)^{3/5}/(m_1+m_2)^{1/5}$), $d_l$ is the luminosity distance of the source to the observer and $\mathcal{I}$ is a function of the angle between angular momentum vector $\vec{L}$ and the line of sight vector $\hat{n}$. 

\section{Gravitational Lensing Authentication Technique using Non-modelled Cross-correlation Searches}\label{basicformalism}

\begin{figure}
    \centering    \includegraphics[width=0.48\textwidth]{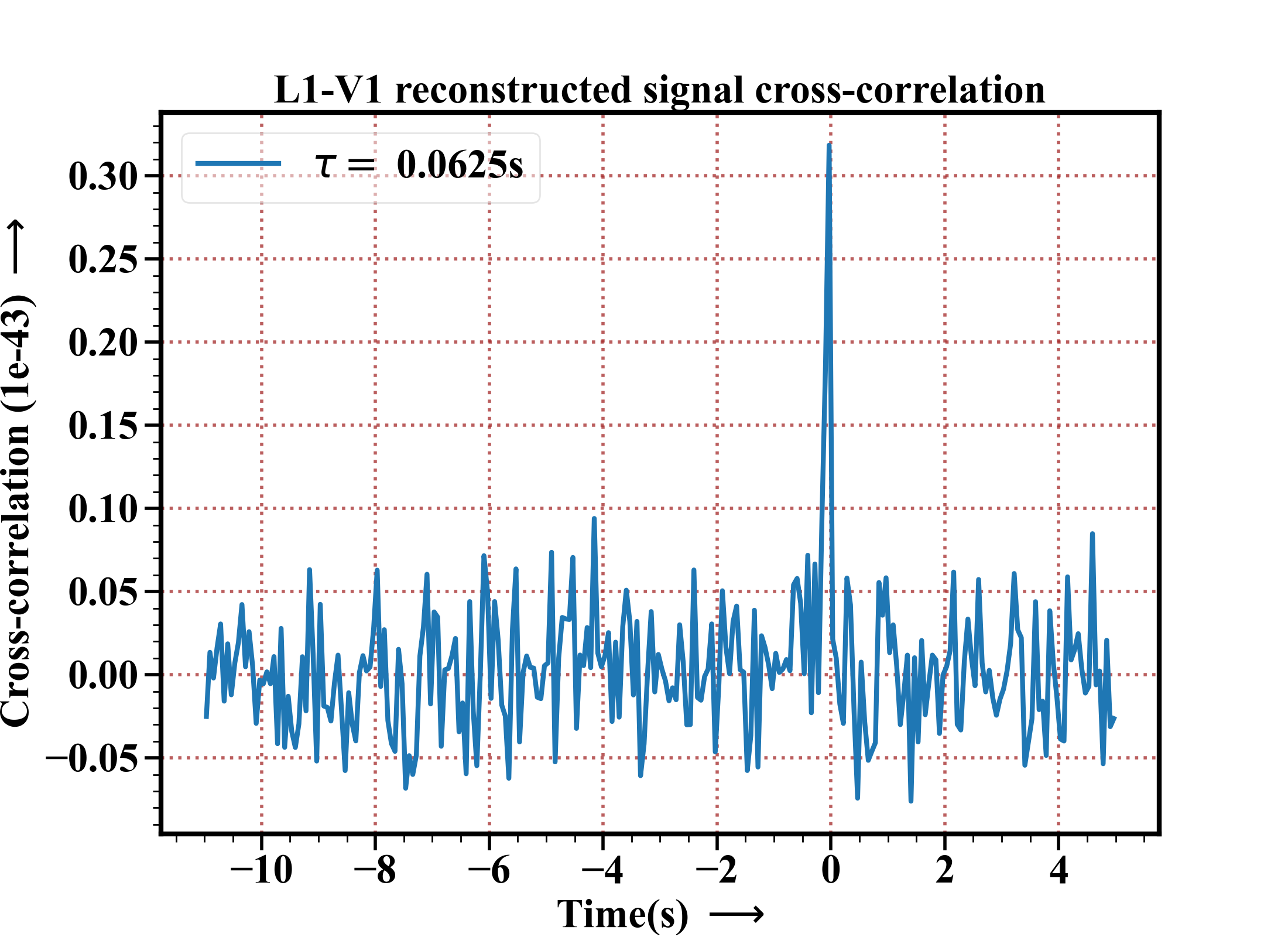}
    \caption{The figure shows the cross-correlation between two reconstructed plus polarization lensed signals. The cross-correlation on the signals shows a monotonic rise followed by a sharp drop after t=0s (the instant when the merger occurs). The cross-correlation is performed from the signals reconstructed from L1 and V1 detectors at two different times.}
    \label{fig6a}
\end{figure}

We claim that a lensed GW signal can be detected by cross-correlation between two time-series data reconstructions containing one polarization of the image signals. For strong lensing cases, multiple images of GWs from the same source are formed by some massive lensing object(s) and images different magnifications $(\sqrt{\mu_i}'s)$ arrive at different times. The time delay between lensed signals can vary between minutes to months for galaxies \citep{PhysRevD.97.023012, 10.1093/mnras/sty2145, 10.1093/mnras/sty411} event up to years for galaxy cluster lens system \citep{Smith_Berry_Bianconi_Farr_Jauzac_Massey_Richard_Robertson_Sharon_Vecchio_etal_2017, 10.1093/mnras/sty031}. Thus for the strong lensing case, the cross-correlation signal can be obtained between the reconstructed data pieces at two different times. 

However, for the micro-lensing case, a single image with a beating pattern envelope is formed, and the cross-correlation needs to be performed with that single image as observed by different detector pairs. 
The application of this technique for micro-lensing events follows a similar but distinct approach. This will be presented in a follow-up paper \citep{Chakraborty:2024:new}. 
To understand the basic theory behind gravitational lensing, we have discussed the some aspects of strong lensing and microlensing in the appendix \ref{app2}. 

\subsection{Mathematical Formalism}

In fig. \ref{fig5} we have shown strongly lensed GW data for three detectors generated by the PYCBC \citep{alex_nitz_2024_10473621}, a PYTHON-module used extensively for gravitational wave detections and simulations. We created the lensed signals as $h_1^L(t) = \sqrt{\mu_1} h(t)$ and $h_2^L(t) = \sqrt{\mu_2} h(t)$. We take stationary, uncorrelated, Gaussian noise generated from the detector noise power spectral density (PSD) of LIGO and VIRGO detectors \footnote{We have used aLIGOAdVO4T1800545 for the LIGO-O4 PSD and AdvVirgo for the VIRGO PSD for the noise generation.}. We have used the phenomenological waveform model IMRPhenomD \citep{Khan:2015jqa} to generate GW signals. The signals have different strengths in the detectors depending upon the antenna pattern of the detector towards the source position in the sky. Thus, the data at any detector $i$ in the time-domain consists of the signal $s_i$ and the noise $n_i$ and is given by, 
\begin{equation}\label{data}
    d_i (t) = s_i (t) + n_i (t) = F_{\times i} h_{\times}(t) + F_{+ i} h_{+}(t) + n_i(t),
\end{equation}
where $F_{\times i}$ and $F_{+ i}$ are the detector antenna patterns of the detector $i$ for the cross and plus polarization respectively. Given that the sky-localization of the event is sufficiently well (90\% credibilty region of around 100 sq. degrees) we can find the detector antenna function $F_{\times i}$ and $F_{+ i}$ for every detector. Therefore a prior knowledge of the antenna pattern towards a sky position for a network of multiple (at least two) non-coaligned detectors, we can uniquely determine the best-fit contribution of the strain for both plus and cross-polarization contaminated by the instrument noise (we have discussed the signal reconstruction procedure in great details in appendix \ref{app:signal_reconstruction}). We use these two noisy polarization signals for the cross-correlation technique for our lensing searches. We define the cross-correlation between two time-domain functions as,
\begin{equation}
C_{12} (t) = f_1 \otimes f_2 = \dfrac{1}{\tau} \int_{t-\tau/2} ^{t+\tau/2} f_1 (t') f_2 (t' + t_d) dt',
\end{equation}
where $t_d$ is a free parameter that is varied to sweep through the domain of the function.

\begin{figure}
  \centering
  \includegraphics[width=0.48\textwidth]{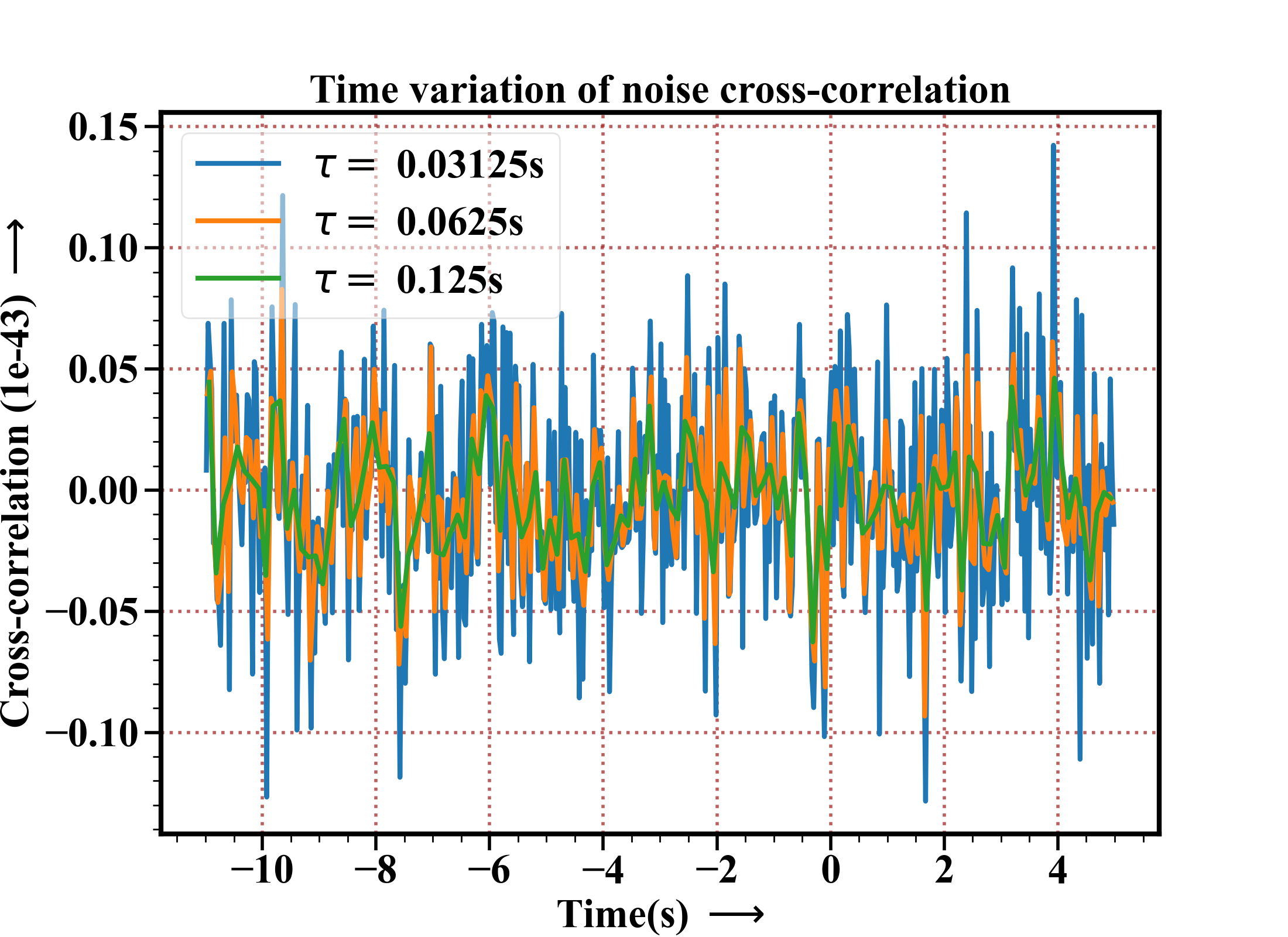}
  \caption{The figure shows the variation of noise cross-correlation with respect to cross-correlation timescale at different instants. It shows random variation with a mean of zero. The fluctuations go down with increasing cross-correlation timescale.}
  \label{fig6}
\end{figure}

When a gravitational wave gets strongly lensed, the amplification factor becomes frequency-independent. 
The phase of the signal is shifted by a frequency-independent constant amount. Therefore the waveform of the signal of the GW gets modulated by frequency-independent amplification and phase factors in strong lensing. Then cross-correlating any two GW signals in proper phase would help us find the similarity between them. Since there is a phase difference of $\pi/2$ between the two polarizations and an antenna pattern function modulates the amplitude of each polarization, it turns out that the cross-correlation between the whole strain data containing both polarizations is not fruitful. Therefore, we cross-correlate between the best-reconstructed contribution on one polarization (say, plus polarization) contaminated by the detector noise. However, the lensed signal may contain a $\pi/2$ phase shift for type-II images \citep{Takahashi_2003}. We then need to cross-correlate between the plus polarization of a signal with the cross polarization of the lensed counterpart. We have discussed how to deal with this case in the appendix \ref{app:phasetest}. In our simulations, we chose to reconstruct the plus polarization signal of the GW signal \footnote{However, cross-polarization is needed to check if there is a phase difference between the two GW image signals e.g. $\pi/2$ phase difference for type-II images.} contaminated by a Gaussian noise generated from the detector noise PSD.
The reconstructed plus polarization signal with a pair detectors (say $i$ and $i'$, let's call it together as $x$) $x$ is of the form, 
\begin{equation}\label{eq:reconstructed_data}
    d^{+}_{x} (t) = h^{+}_{x}(t) + n^{+}_{x}(t).
\end{equation}
Then the cross-correlation between best-reconstructed plus polarization signals by different pairs of detectors at two different times (given by $d^+_x$ and $d^+_{x'}$) can be defined as,\footnote{Similarly, $d^+_{x'}$ is developed using other two detector combination, let's say $i'$ and $i''$ . }
\begin{align}\label{eq7}
    D_{x x'} (t) &= d_x^+ \otimes d_{x'}^+ = \frac{1}{\tau} \int_{t-\tau/2} ^{t+\tau/2} d^+_{x} (t') d^+_{x'} (t' + t_d) dt', \nonumber\\
    &= S_{x x'}(t) + N_{x x'}(t) + P_{x x'}(t) + Q_{x x'}(t), \nonumber\\
    & \approx S_{x x'}(t) + N_{x x'}(t) .
\end{align}
The equation of the second expression contains four terms: $h^{+}_x \otimes h^{+}_{x'}$, $n^+_{x} \otimes n^+_{x'}$, $h^{+}_x \otimes n^+_{x'}$, $n^+_{x} \otimes h^{+}_{x'}$ respectively, where `$\otimes$' denotes the cross-correlation between them \footnote{This assumes that the choice of $t_d$ makes the cross-correlation between the two signals in phase. In practice, $t_d$ is a free parameter, while performing a search we have to vary it.}. The cross-correlation is performed over a timescale of $\tau$, with $\tau$ chosen to be a few tens of cycle duration (typically $\mathcal{O}(10^{-2} s)$), the cross-terms $P_{xx'}$ and $Q_{xx'}$ tend towards zero. In this approximation, we are left with only two terms $S_{xx'}$ and $N_{xx'}$ containing the signals' cross-correlation and the noises' cross-correlation \footnote{The term $N_{xx'}$ also tends to zero when the noise in the two data are uncorrelated, else will be non-zero.}.

\begin{figure}
  \centering
  \includegraphics[width=0.48\textwidth]{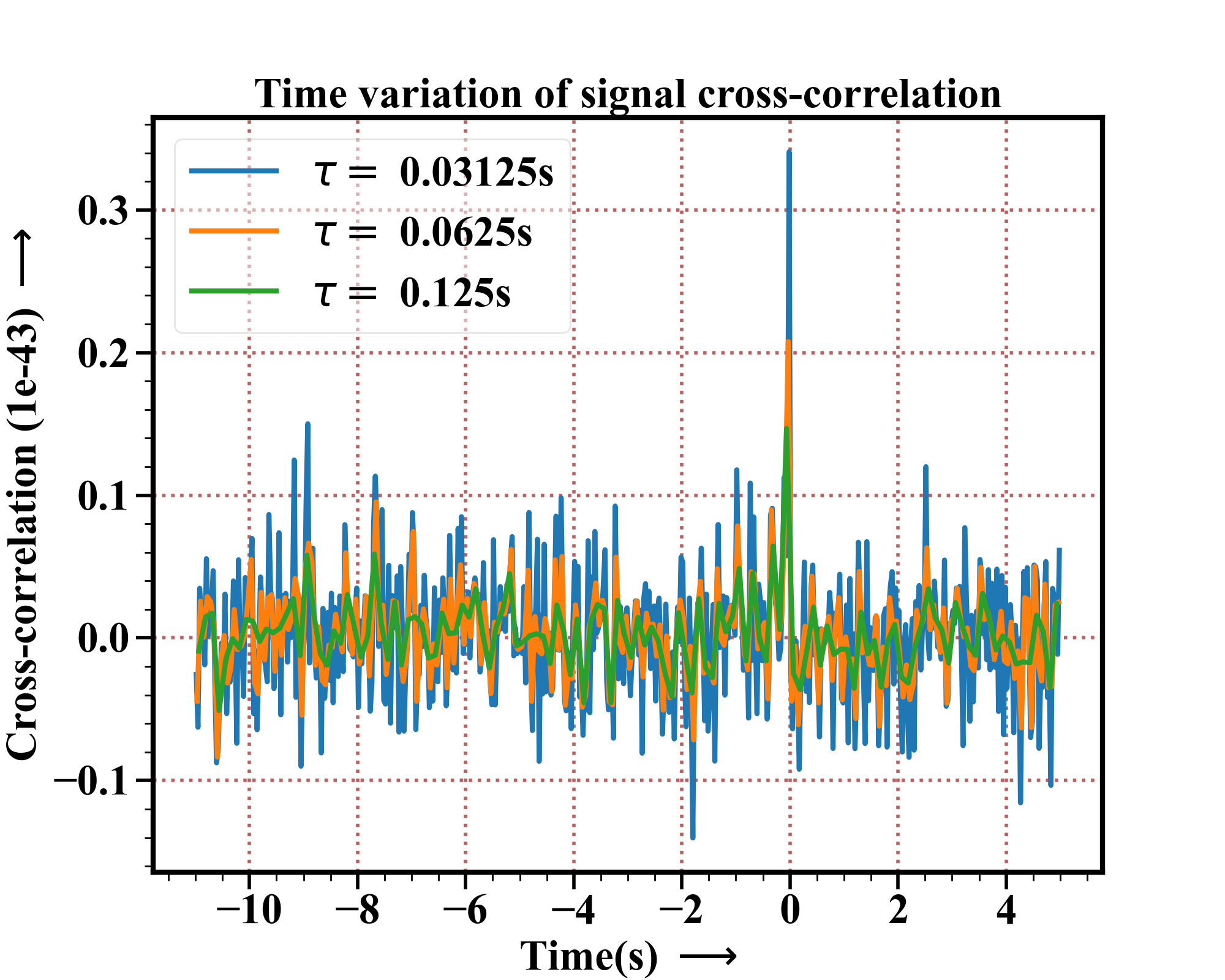}
  \caption{The figure shows the variation of cross-correlation on the strongly lensed signals with cross-correlation timescale. With increasing cross-correlation timescale, the noise fluctuations go down making the signal cross-correlation more prominent.}
  \label{fig7}
\end{figure}

Fig. \ref{fig6a} shows the cross-correlation performed when the signals are in phase. The merger happens in the signal at $t=0s$ therefore we observe a bump in the cross-correlation near that time. The cross-correlation timescale $\tau$ can be chosen accordance with the duration of the time-domain signal, it can be chosen in such a way that it is longer than the typical cycle duration of the GW but much shorter than the whole of the signal duration. The time difference between the data products, $t_d$, for a blind lensing search is a free parameter that can be varied from seconds to years. 
We can define the cross-correlation of reconstructed signals from the same detector pair but with a time delay of $t_d$ accordingly,
\begin{equation}
    D_{x x} (t) = \frac{1}{\tau} \int_{t-\tau/2} ^{t+\tau/2} d^+_{x} (t') d^+_{x} (t' + t_d) dt' \approx S_{x x}(t) + N_{x x}(t) .
\end{equation}

Fig. \ref{fig6} shows the cross-correlation of noise at two different times when the signal is not present. The random noise fluctuations have a mean of zero and thus the cross-correlation of two noise entities for a long enough timescale, always tend towards zero. In fig. \ref{fig7} we show the cross-correlation between the GW plus polarization signals with variation in the cross-correlation timescale. The signal has its max amplitude at t=0s, thus we can observe a gradual rise in the cross-correlation signal followed by a peak around t=0s followed by a drastic fall. Then it is again dominated by noise fluctuations. 
As the cross-correlation timescale is increased, the random noise fluctuations tend towards zero making the cross-correlation peak more prominent. 
Thus if we cross-correlate two similar signals in phase, we expect to observe a positive, monotonically increasing cross-correlation as the chirping nature of the GW continues, followed by a sharp drop after the instance of the BBH merger. In appendix \ref{app:phasetest}, we have shown a few tests to show the robustness of this technique in differentiating between cross-correlation of two signals from the cross-correlation of a signal with some noise. 

\begin{figure}
    \centering
    \includegraphics[width=0.48\textwidth]{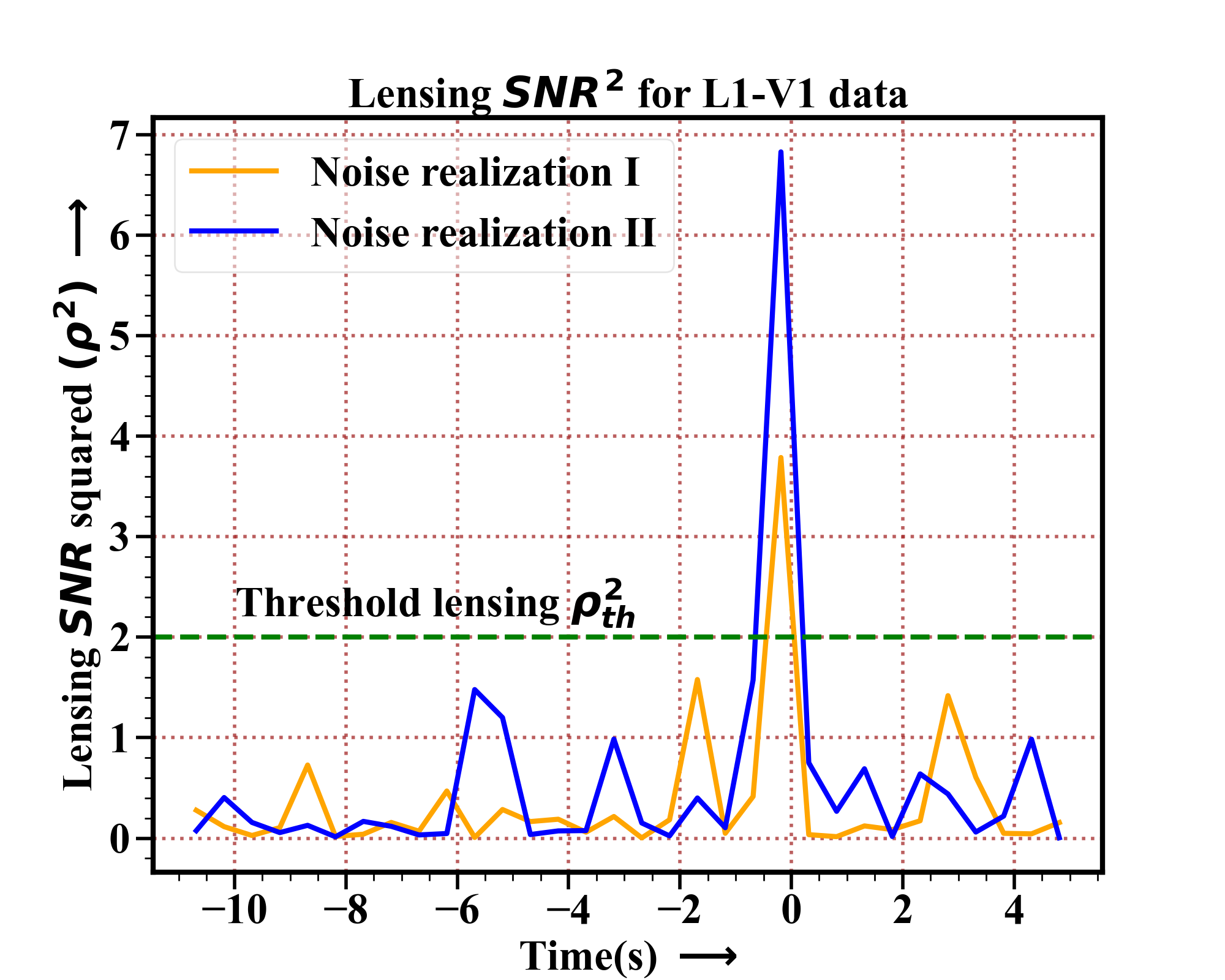}
    \caption{The figure shows the lensing signal-to-noise ratio for the strong lensing case, with $\tau_{snr} =0.5s$ with $\tau_{cc} = 0.0625s$. It depicts the typical deviation of the reconstructed signal cross-correlation from the noise cross-correlation. By setting a threshold lensing SNR, we obtain information about the significance of the event being lensed.}
    \label{fig11}
\end{figure}

Sometimes, there can be persistent noise sources near a detector site. In that case, the noise at two different instances become correlated and we cannot take the signals reconstructed from that detector. Then the cross-correlation has to be performed with reconstructed data chunks from some other pair of detector combination. This is required to avoid the dominance by the $N_{xx'}$ term over the $S_{xx'}$ term. We claim that a blind search in all available LVK data as well as data from the next-gen detectors has to be performed in search of lensed GW signals using this technique.

To generalize, if there are $n$ images created by strong lensing, then with a network of $m$ detectors, we can have at most $ \frac{mn(mn-1)}{2} $ number of cross-correlations \footnote{We assume that all those signals are well separable in time such that they have no overlap.}. However, $n \frac{m(m-1)}{2}$ of these are the cross-correlations between the same signal from different detector pairs, and for obvious reasons they do not contain any important information for the lensing analysis.

\begin{figure}
    \centering
    \includegraphics[width=0.48\textwidth]{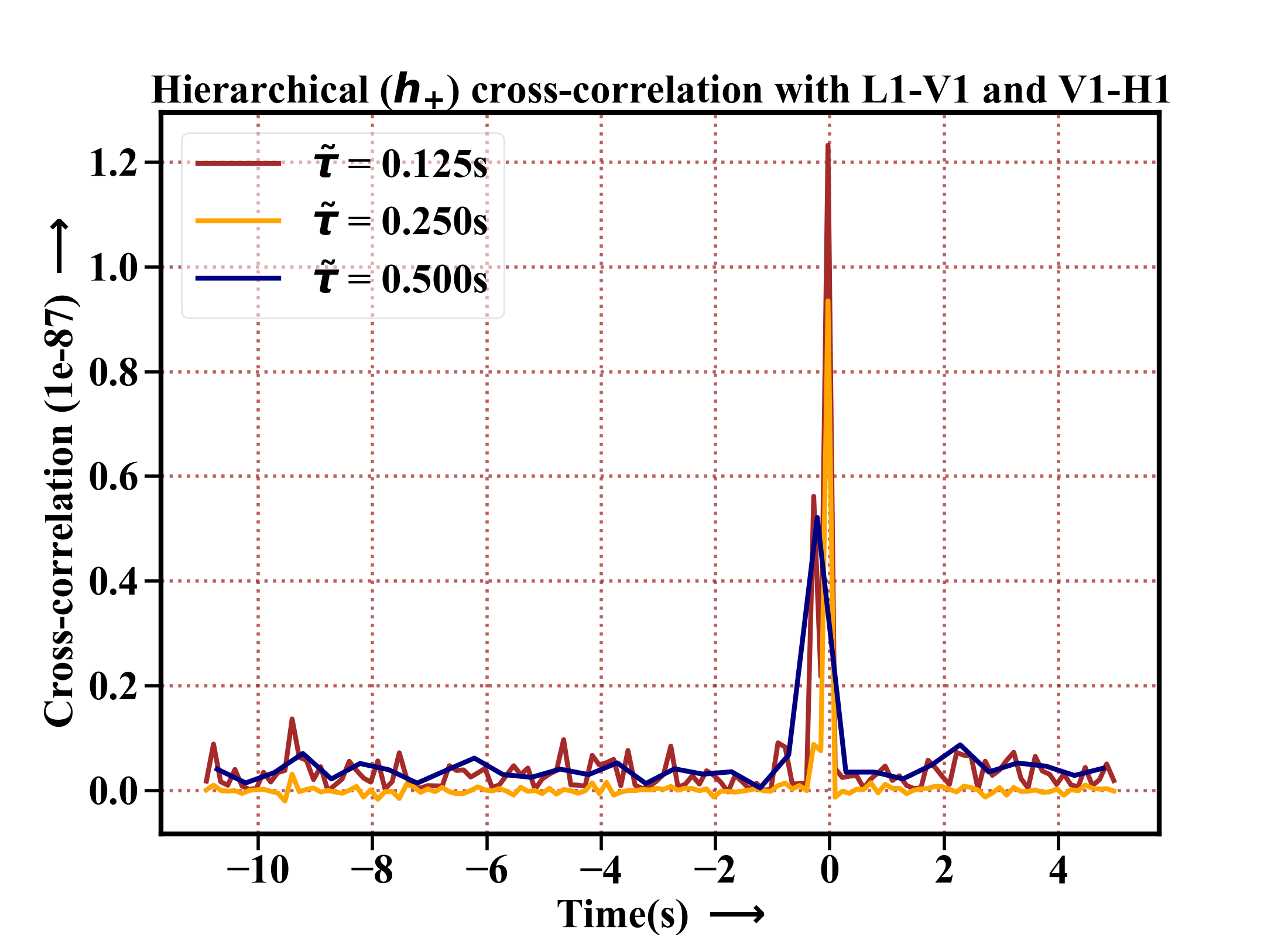}
    \caption{The figure shows the hierarchical cross-correlation performed on two cross-correlation signals. The noise is further suppressed than the reconstructed signal cross-correlation case. However, any two positive noise cross-correlations (positive over some time) can, in principle, produce a similar hierarchical cross-correlation pattern. However, the phase information of the signal is not used here. This means depending on hierarchical cross-correlation alone should be made with proper caution.}
    \label{fig:hierarchicalcross}
\end{figure}

\subsection{Confirmation of detection of a lensed signal}

We observed the cross-correlation signal when two signals are present and the cross-correlation is performed in right phase. However, the confirmation of lensing requires mitigating the noise as much as possible. To qualify an event as a lensing candidate, we take the binned average of these reconstructed signal cross-correlations $\langle D_{xx'}(t)\rangle$ and calculate how much it deviates from typical bin-averaged noise cross-correlation $\langle N_{xx'}(t + \Delta t)\rangle$. The noise cross-correlation ($N_{xx'}(t + \Delta t)$) is calculated when no signal is present in the data. The deviation of the averaged data cross-correlation from the averaged noise cross-correlation helps us calculate an effective lensing signal-to-noise ratio or SNR ($\rho$). 
We define the square of lensing SNR as:
\begin{equation}\label{eq9}
    \rho ^2(t) = \left ( \frac{\langle D_{xx'}(t)\rangle-\langle N_{xx'}(t+\Delta t)\rangle}{\sigma_N (t+ \Delta t)} \right )^2 ,
\end{equation}
where $t + \Delta t$ is a chosen time when there are no signals around \footnote{For all those instant considered, the match-filtering SNR is below 4.} and we calculate the noise cross-correlation $N_{xx'}(t + \Delta t)$ at this point of time; ${\sigma_N (t+\Delta t)}$ being the standard deviation of the $N_{xx'}(t + \Delta t)$ in the chosen bin.
Given the timescale $\tau_{snr}$ corresponding to n points of cross-correlation signal, the average is taken as, 
\begin{equation}
    \langle D_{xx'}(t)\rangle = \frac{\sum _{i=1} ^{n}  D_{xx'}|_{i}}{n},
\end{equation}
where $D_{xx'}| _{i=1} = D_{xx'}(t-\frac{\tau_{snr}}{2})$ and $D_{xx'}|_ {i=n} = D_{xx'}(t+\frac{\tau_{snr}}{2}
)$ and the same applies for $\langle N_{xx'} \rangle$.

The definition of the match-filtering SNR (that is used to confirm the presence of GW signals in data) = $4 \Re{(\int_0 ^\infty \frac{h^{*}(f)d(f)}{S_n (f)})}$ , where $d(f)$ be frequency domain data, $h^*(f)$ be the complex conjugate of the frequency domain GW signal template and $S_n(f)$ is the power spectral density of the noise for the particular detector. Notably, the definition for lensing SNR is quite different from the match-filtering SNR. We require at least two signals for calculating lensing SNR as compared to the requirement of one signal for calculating match-filtering SNR. However, both of them provide us with an idea how confident we are with our detection. We validate the candidature of the event(s) being lensed by calculating the lensing SNR of the event(s). Events that are above the threshold ($\rho_{\rm event} \geq \rho_{\rm th}$) lensing SNR are referred to as candidates for strong lensing, given their sky positions have some overlap.

\begin{figure*}
    \centering
    \includegraphics[width=0.9\textwidth]{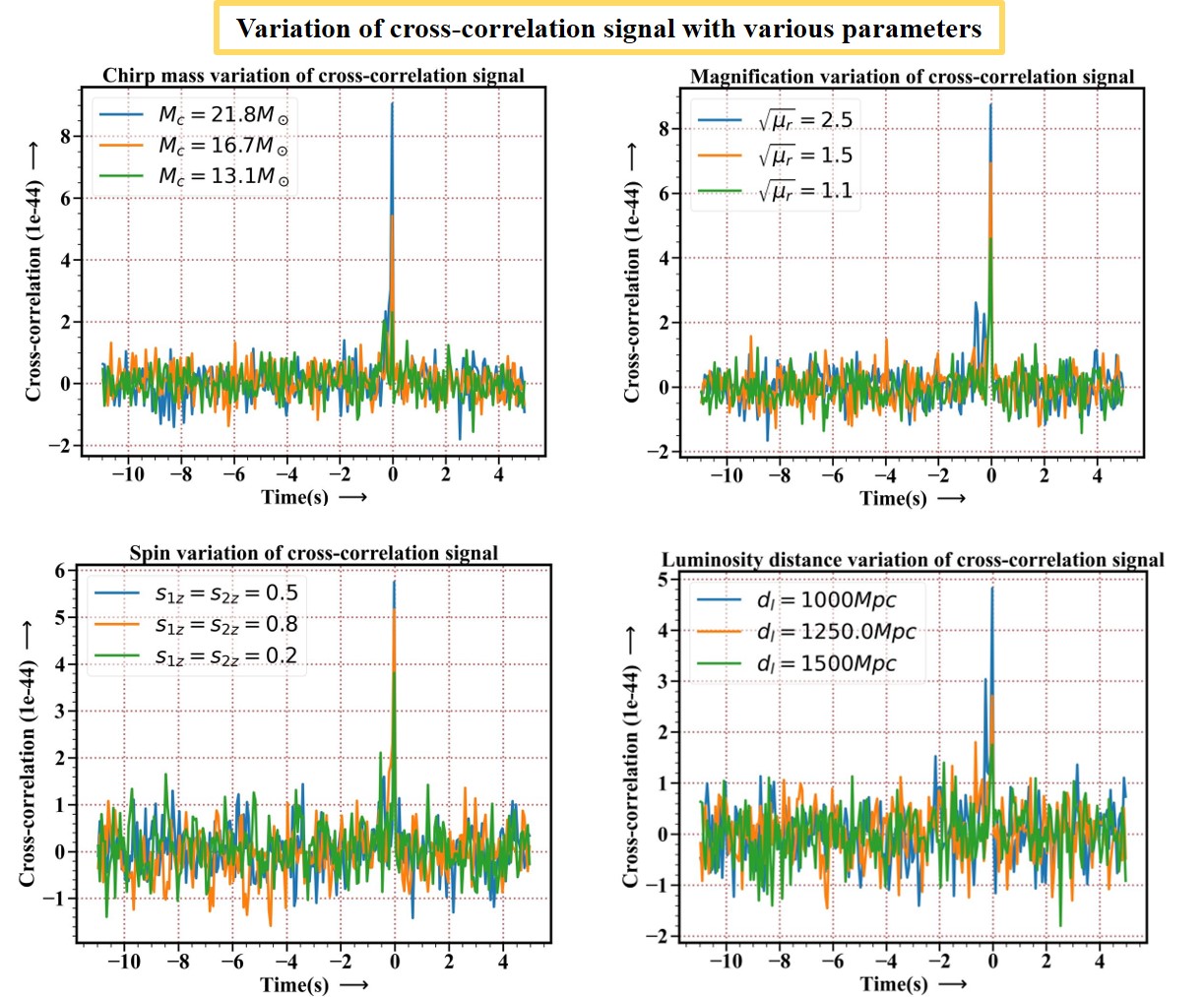}
    \caption{The figure shows the variation of the cross-correlation signal with respect to the source parameters (chirp mass of the source, its luminosity distance, and the spin components along its orbital angular momentum) and lens-induced parameters (magnification). The timescale of the cross-correlation is chosen to be $\tau = 0.0625s$. The cross-correlation peak strength gives us an idea about the possible upper bounds in different source and lens parameters for which \texttt{GLANCE} will work properly.}
    \label{fig12}
\end{figure*}

Fig. \ref{fig11} shows the lensing SNR for two similar GW signals at different instants. We have plotted the lensing SNR for two different noise realizations. There is an observable jump in the lensing $SNR^2 (\rho^2)$ around $t=0s$ because that's the instance when two signal amplitudes are maximum. The threshold value of $\rho^2$ above which we call a lensing candidate can be calibrated by observing the typical SNR of noise cross-correlation. Apart from that peak, we can also observe that for both of the noise realizations, we have several other spikes. Again these spike heights decease as we increase the lensing SNR timescale. Judging by the typical SNR of noise cross-correlation for these two scenarios (which shows a significant drop in the lensing SNR for some unfavourable noise realizations) we can choose this threshold to be at $\rho^2 _{th} = 4$ for the case with $\tau_{cc} = 0.03125s$ and $\tau_{snr} =0.5s$ \footnote{The threshold lensing SNR is dependent on the cross-correlation timescale and the lensing SNR timescale for reasons discussed already. However, such choices need not to be universal, but rather be set according to the noise specifications of the detectors.}. The timescale of SNR can be made as large as the duration of the signal, for maximum suppression of the noise fluctuations. Even then, there can be peaks in the SNR of the same order/higher than the signal SNR when there is a correlated strong noise in some detector(s). We have shown the impact of different noise realizations on the lensing SNR in appendix \ref{app:noise}. The validation of lensing (and distinguishing it from some noise artifacts) can only be made possible when multiple cross-correlations performed between different reconstructed signals show a significant bump in lensing SNR at the same instant \footnote{With the inclusion of LIGO-India, the polarized signal reconstruction will improve significantly.}. 

There can be cases when the lensing SNR for the signals is not distinct from the noise fluctuation SNRs, however a cross-correlation signal seems to be present. Then all reconstructed signals, that show a similar cross-correlation pattern, can be investigated further to the strain level data to check whether any known noise characteristics are present or not. Also, using our current understanding of the amplitude and the phase of the GW, we can estimate the source parameters jointly from on those two strains to confirm convergence of the source parameters and if no convergence is observed, we do not proceed further with those strain pieces on the lensing analysis since the strains do not contain a similar characteristic signal.

In the hope of further improving the lensing detection significance, we used hierarchical cross-correlation $\Delta_{x x'}$ performed between different two different cross-correlation signals $D_{xx'}$ . The second-order cross-correlation takes the cross-correlation of the data and cross-correlates it again; it is defined as,
\begin{equation}
    \Delta_{x x'} (t) = \frac{1}{\tilde{\tau}} \int_{t-\tilde{\tau}/2} ^{t+\tilde{\tau}/2} D_{xx} (t') D_{x'x'} (t' + T_d) dt',
\end{equation}
where $\tilde{\tau}$ is the cross-correlation timescale here and $T_d$ is the time-shift (a free parameter) required for the cross-correlation signals to overlap. In fig. \ref{fig:hierarchicalcross} we can observe that the 2nd-order signal cross-correlation is more prominent than the 1st-order cross-correlation signal taken between the noisy signals with one polarization. Noise fluctuations are suppressed with increasing $\tilde{\tau}$. However, the 2nd order cross-correlation does not incorporate the phase of individual signals, any two gradually rising functions both of positive (or negative) values of any origin, can in principle, produce a 2nd order cross-correlation similar to this. We emphasize that the higher-order cross-correlation works better if the signal is pronounced for long duration (minutes-long GW signals, so that finding a similar pattern between the signal cross-correlations is easier), but for short signals (of the order of seconds), such higher-order correlations alone are not reliable in terms of searching for lensed signals. Thus, to avoid lensing triggers that are generated from some strong, correlated and short noise source, we would rely more on the in phase cross-correlation and its lensing SNR.

\subsection{Dependency of Cross-correlation signal on source and lens parameters}

\begin{figure*}
    \includegraphics[width=0.9\textwidth]{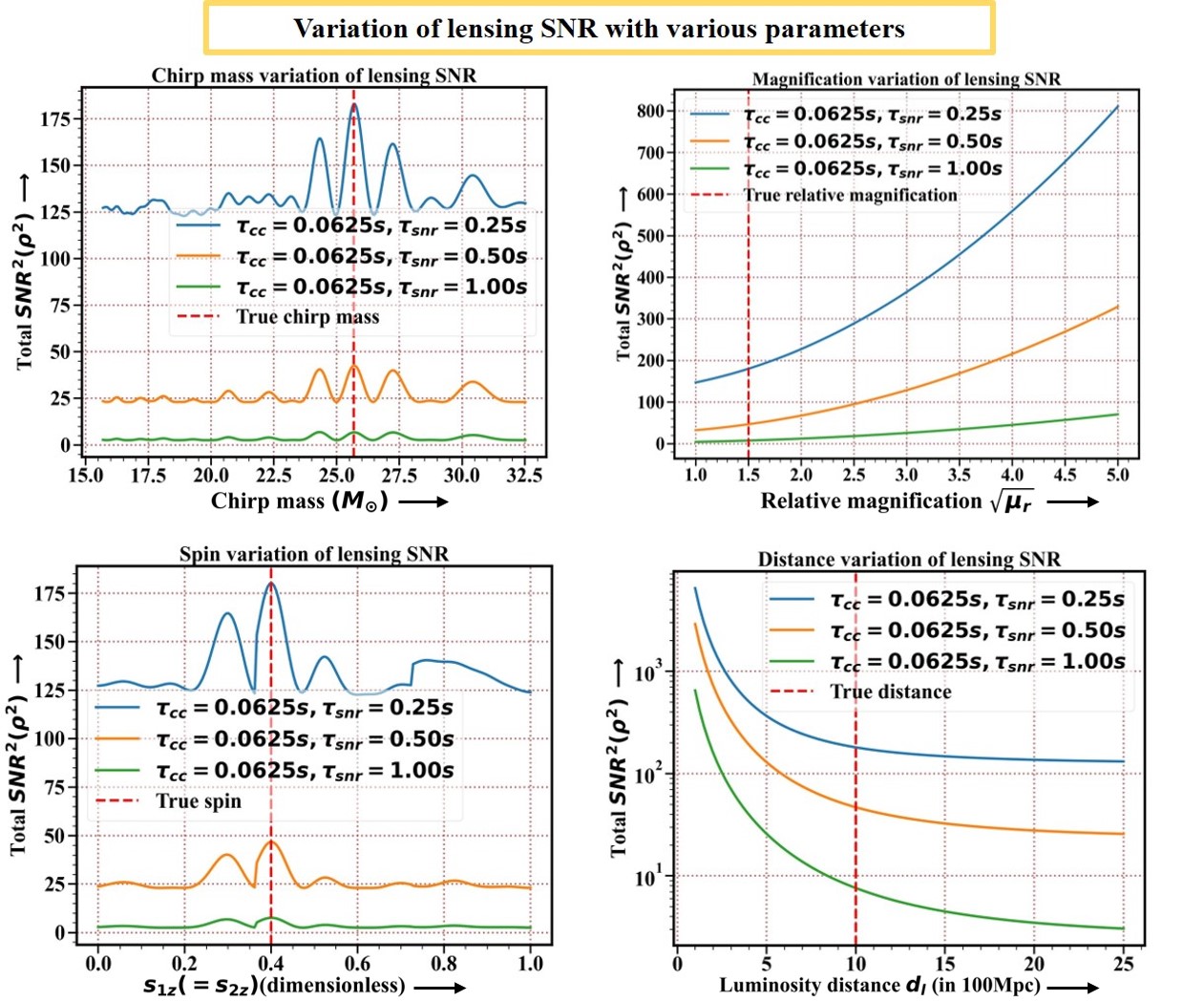}
    \caption{The figure shows the variation of the lensing SNR with chirp mass, relative magnification, luminosity distance, and spins. It shows that the recovery of an injection parameter is almost successful (when other parameters are kept unchanged) for chirp masses and spins, but not for luminosity distance and relative magnification. We have also shown the variation of lensing SNR on the SNR timescale $(\tau_{\rm snr})$. We can observe many small peaks around the true injection values which die down as we increase the SNR timescale. However, there are peaks around the true injection for chirp mass and spins. This indicates that there lies some chances of false lensing triggers even when the parameters of the two signals do not match.}
    \label{fig18}
\end{figure*}

\begin{figure*}
    \centering
    \includegraphics[width=0.9\textwidth]{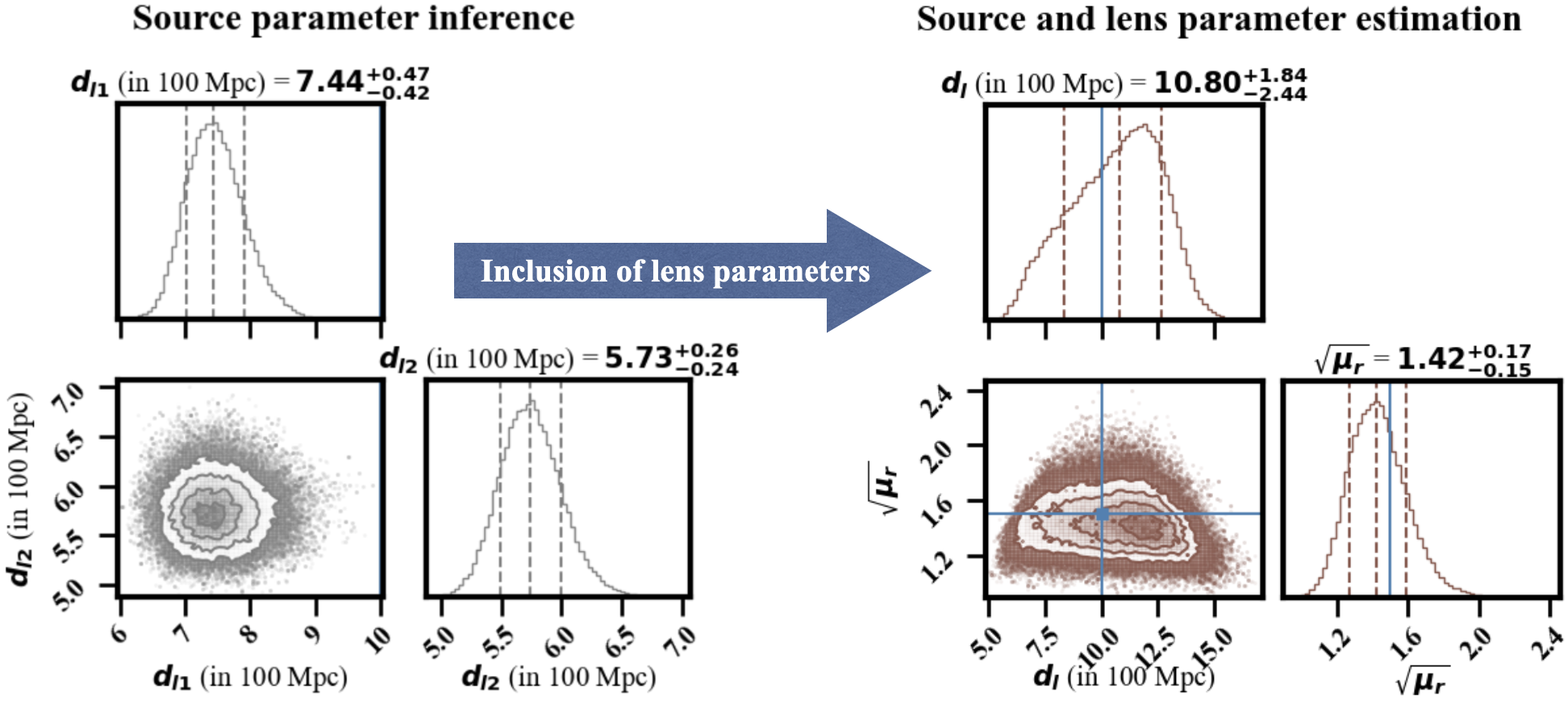}
    \caption{The figure shows that the parameter space for no lensing hypothesis is different from that considering lensing. Without the strong lensing hypothesis, there is no existence of lensing magnification which plays a role in inferring the true luminosity distance of the source. It shows that with no lensing hypothesis, if two strongly lensed images have a relative magnification of $\sqrt{\mu_r}=1.5$ (with individual magnifications as $\sqrt{\mu_1}=1.2$ and $\sqrt{\mu_2}=1.8$) with sources at $d_l= 1$Gpc with BBH component masses of $m_1=15M_{\odot}$ and $m_2=25M_{\odot}$, we get a wrong inference of individual luminosity distance of the two events, making them appear like two separate events. However, with the lensing hypothesis, the inclusion of lens parameters along with the source parameters helps us characterize the source and the lens jointly making a correct parameter estimation for the lensed events. The injected parameters luminosity distance ($d_l$) and relative magnification ($\sqrt{\mu_r}$) are well recovered in our inference of the 2D joint source and lens parameter space.}
    \label{fig:infer_prob}
\end{figure*}

It is necessary to check the source parameter dependency of the signal cross-correlation to understand the possible range of parameter space that can explored using the cross-correlation method. Thus in fig. \ref{fig12}, we have shown the dependency of the cross-correlation signal on different source parameters as well as modification by the lens, i.e. magnification, and observe the signal cross-correlation strength as compared to noise cross-correlation. We choose a fixed set of parameters $(m_1, m_2, d_l, s_{1z}, s_{2z}, \iota, \sqrt{\mu_1}, \sqrt{\mu_2}) = (25M_{\odot}, 15M_{\odot}, 1 {\rm Gpc}, 0.2, 0.1, 0.2, 1, 1.5)$. We then varied one of the parameters to observe how the cross-correlation signal varies. The first of the plots shows the variation of the signal cross-correlation with the chirp mass (by varying $m_1$ and $m_2$ both). As we have increased the chirp mass of the source, individual signals get stronger resulting in a higher peak value in the cross-correlation. As can be seen, the method can work well for sources with chirp mass as low as 16.7 M$_{\odot}$. The next plot shows the variation of cross-correlation signal with respect to the relative magnification $\sqrt{\mu_r} = \sqrt{\mu_2 /\mu_1}$ (we varied $\sqrt{\mu_2}$  keeping $\sqrt{\mu_1}$ fixed). Relative magnification signifies the boost in the signal because of the presence of the lens. As relative magnification goes up, signal has a greater contribution in the data in comparison to noise, so the signal cross-correlation peak goes up with increasing magnification. It is observed that, even with relative magnification of 1.1, the cross-correlation signal peak can be resolved. The next plot shows the variation of the cross-correlation signal with respect to the spins of the black holes (along the orbital angular momentum direction). We see some variation but cannot conclude on the trend of the variation here. The final plot shows the variation of the cross-correlation signal with respect to the luminosity distance. Increasing the luminosity distance makes the signal fainter, thus with increasing distance cross-correlation peak height goes down. As can be seen, up to a luminosity distance of 1.25 Gpc, the cross-correlation peak is barely resolvable. As seen here, magnification and luminosity distance are two counter-acting quantities affecting the strength of the signal. We will later see, there lies a degeneracy while estimating these two parameters, which we will resolve in a later section.

In fig. \ref{fig11}, we found how the lensing SNR ($\rho$) appears at different instants as the signal arrives. We added the values of $\rho^2$ at all the instants around the signal lensing SNR to find the total $\rho^2$. We took a fixed set of source parameters and magnification to create our data-I given by $(m_1, m_2, d_l, s_{1z}, s_{2z}, \iota, \sqrt{\mu_1}, \sqrt{\mu_2}) = (35M_{\odot}, 25M_{\odot}, 1 {\rm Gpc}, 0.4, 0.4, 0.2, 1, 1.5)$. We then fixed all parameters except one to create the data-II. Then we varied that single parameter and checked how the lensing SNR varies as the variable parameter approaches the data-I injection parameter. 
Variation of the total $\rho^2$ provides us with an idea about how correctly we can estimate a parameter and how the source parameters are degenerate with the lens parameters. 

We show the variations of the total of squared SNR vs different parameters in fig. \ref{fig18}. In the first plot, we varied the chirp mass (by fixing $m_1$ and varying $m_2$) and observe that the total lensing SNR peaks at the injected chirp mass in data-I. However, there are some other smaller peaks at some nearby chirp masses. This indicates that there is some overlap of the two signals even if their source chirp masses are different. This is a case of false triggers in this lensing analysis, that we will discuss in brief in a later section. In the second plot, we vary the luminosity distance. The lensing SNR goes monotonically down with increasing source luminosity distance and we do not observe any peak at the injected value of data-I. This is expected because the signal become fainter, the lensing SNR should go down. In the third plot, we vary the spin component along the orbital angular momentum. We can see that, there is a peak at the injection value as we keep varying the spin. However, like before, there are some other smaller peaks at values other than the injection hinting at the possibility of raising false triggers for those spin values. In the fourth plot, we varied the relative magnification $\sqrt{\mu_r} = \sqrt{\mu_2 /\mu_1}$ (we varied $\sqrt{\mu_2}$ keeping $\sqrt{\mu_1}$ fixed). We do not see any peaks at the injection as expected, because the total lensing SNR goes up monotonically with increasing $\sqrt{\mu_r}$. 

Therefore, for the mass and the spin case, as we approach the values of the data-I injected parameters by varying data-II variable parameters, we observe a peak which diminishes as we move away from the injection value. This shows that, although a background of total $\rho^2$ is always present, it only peaks when we are significantly close to the injection. Only for those cases, the signals overlap nicely resulting in a peak lensing SNR. However, there are possible ranges on chirp mass and spin over which the lensing SNR is still considerable. This hints at the existence of false triggers in lensing searches. The plots also show that, as we increase the SNR timescale $\tau_{snr}$, the smaller lensing SNR fluctuations smooths out and the strong peaks remain. Certain unfavourable noise realizations and bad choice of timescales \footnote{$\tau_{snr}$ should be much longer than the cross-correlation timescale, now called as $\tau_{cc}$ to avoid any confusion} can lead to the inability of the recovery of the injected parameter. Another way to quantify the SNR of the signal is to pick the maximum SNR around the time a signal is present i.e the cross-correlation signal peaks, instead of using summed SNR of all instants around the signal. The performance of our method using max SNR is shown in appendix \ref{app:choicesnr}.

However, we also observe that the variation of the total $\rho^2$ with respect to relative magnification or luminosity distance is rather monotonic and it does not show any peak in the lensing SNR. Therefore, one can easily understand that the signal contains degeneracy in the magnification and luminosity distance. Therefore, we need to find a clever way in our parameter estimation techniques to explore the range of values these source and lens parameters can take jointly. We will discuss this thoroughly in the coming section.

\section{Lens and source characterization of the strong lensing candidates} \label{section6}

Since gravitational lensing substantially modifies the waveform of gravitational waves, inference of the source parameters with no lensing hypothesis produces significantly different results than their inference with lensing hypothesis. This happens because some source parameters with no lensing hypothesis are degenerate with some other source parameters under the lensing hypothesis. This makes source parameter estimation extremely challenging with the lensing in consideration. For strong lensing cases, the lensing modifications do not alter the inference of the masses and spins. Thus the masses and spins inferences from the lensed signal are unaffected by the strong lensing biases. Therefore, the amplification factor in strong lensing is not degenerate with any of those parameters.
However, the signal is amplified by a constant magnification factor thus, only luminosity distance $(d_l)$ and magnification factor $(\sqrt{\mu})$ are such degenerate parameters. This is because the GW strain is inversely proportional to $d_l$ and if lensed, the two signal strains gets a proportionality factor of $\sqrt{\mu_1}$ and $\sqrt{\mu_2}$ respectively. So if lensing is not considered, we would infer the luminosity distances to be $d_l/ \sqrt{\mu_1}$ and $d_l/ \sqrt{\mu_2}$ respectively, making the events identified as two different astrophysical events. Therefore, strong lensing biases our inference of the true source parameters, particularly the luminosity distance of the source. This is shown in fig. \ref{fig:infer_prob} that if two events are identified to be lensed candidates then it is important to identify them as a single astrophysical event and re-estimate their luminosity distance and relative magnifications by combining the two inferred 'apparent' luminosity distances. Even though the error on distance will be larger, it gives a broad, but not an incorrect true distance. Hence makes it possible to count it correctly as an astrophysical event (and not as a two individual events at different incorrect distances).

However, for two data strains, we cannot estimate three degenerate parameters ($\sqrt{\mu_1}$ and $\sqrt{\mu_2}$: the magnification of each signal and $d_l$: the true luminosity distance of the source) from it. Thus the reduced form of the two individual magnifications $\sqrt{\mu_r} = \sqrt{\mu_2/\mu_1}$ along with the luminosity distance $d_l$ are two parameters that we aim to estimate. We took a merging BBH as a transient GW source; the component masses are $25M_{\odot}$ and $15M_{\odot}$ at a distance of 1Gpc and one signal had a magnification of $\sqrt{\mu_1}=1.2$ with the other one having magnification of $\sqrt{\mu_2}=1.8$, so there is a relative magnification of $\sqrt{\mu_r}=1.5$ between the two signals.

However, for microlensing cases masses and spins parameters can be degenerate with the lens-induced parameters on the signal. We are working on our future work to explore the joint parameter space of the source and the lens for the microlensing case \citep{Chakraborty:2024:new}.

Here, for the strong lensing degenerate parameter estimation, we have used Bayesian analysis to estimate the source parameter ($d_l$) and lens modification ($\sqrt{\mu_r}$) on the strongly lensed GW signal. The basic principle of the Bayesian inference technique is the Bayes' theorem which is given by,
\begin{equation}
    p(\vec{\theta} | d) \propto L(d| \vec{\theta}) \Pi(\vec{\theta}),
\end{equation}
where $\Pi(\vec{\theta})$ is the prior on the luminosity distance and relative magnification, provides the prior range of possible values the different source and lens parameters can take. $L(d| \vec{\theta})$ is the likelihood, describing the probability of obtaining the data with the chosen prior for different parameters and $p(\vec{\theta}|d)$ is the obtained posterior for the chosen parameters. There is an overall normalization factor in the denominator, $p(d)$ known as the evidence, does not bear significant importance in this analysis.

The data at any detector $\textnormal{d}_i = F_{\times i} h_{\times} + F_{+i}h_{+} + n_i$ is given in terms of the unlensed model is given by,
\begin{equation}\label{eq:data}
\textnormal{h}^{ul}_i = F_{\times i} h^{ul}_{\times}(\vec{\theta})  + F_{+ i}h^{ul}_{+}(\vec{\theta}) \quad .
\end{equation}

We choose a flat prior for the parameters and a Gaussian likelihood, and obviously we expect to get a Gaussian posterior. The choice of the likelihood for strong lensing is mentioned below:
\begin{equation}\label{eq:likelihood}
\begin{split}
    -\log(L) &= \frac{(d_{1} - h^{l}_1 (\sqrt{\mu_1}, d_l))^2}{2\sigma_{n1}^2} + \frac{\log(2 \pi \sigma_{n1}^2)}{2} \\
    &+ \frac{(d_{2} - h^{l}_2 (\sqrt{\mu_2}, d_l))^2}{2\sigma_{n2}^2} + \frac{\log(2 \pi \sigma_{n2}^2)}{2},
\end{split}
\end{equation}
$d_1$ and $d_2$ are the data chunks that pass through all aforesaid tests, $\sigma_{n1}$ and $\sigma_{n2}$ are the standard deviation associated with the data in the absence of any signals. 


\begin{figure}
    \centering
    \includegraphics[width=0.48\textwidth]{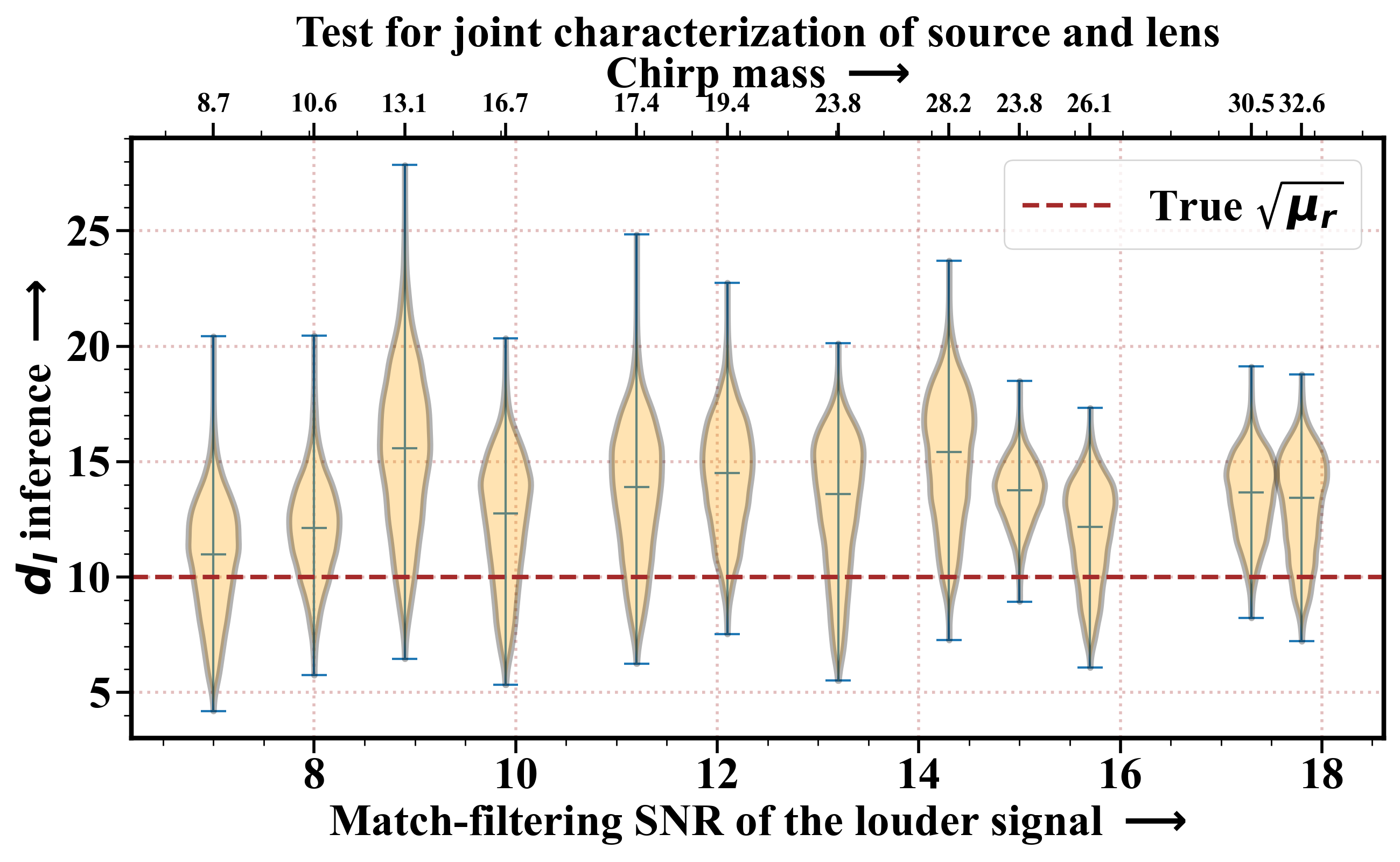}
    \caption{The figure shows the luminosity distance inference for a range of chirp masses which can be translated in terms of match-filtered SNR (since all other parameters are kept fixed). Thus we have shown how the parameter estimation for $d_l$ works while varying the strength of the signal. Thus we have both chirp mass and match-filtered SNR along the x-axis. The figure shows that with increasing match-filtered SNR of the louder signal (which has a relative magnification of 1.5 with respect to the fainter one), the estimation of $d_l$ gets better, and we obtain posteriors with smaller variance around the injection value.}
    \label{fig:dist_infer}
\end{figure}

\begin{figure}
    \centering
    \includegraphics[width=0.48\textwidth]{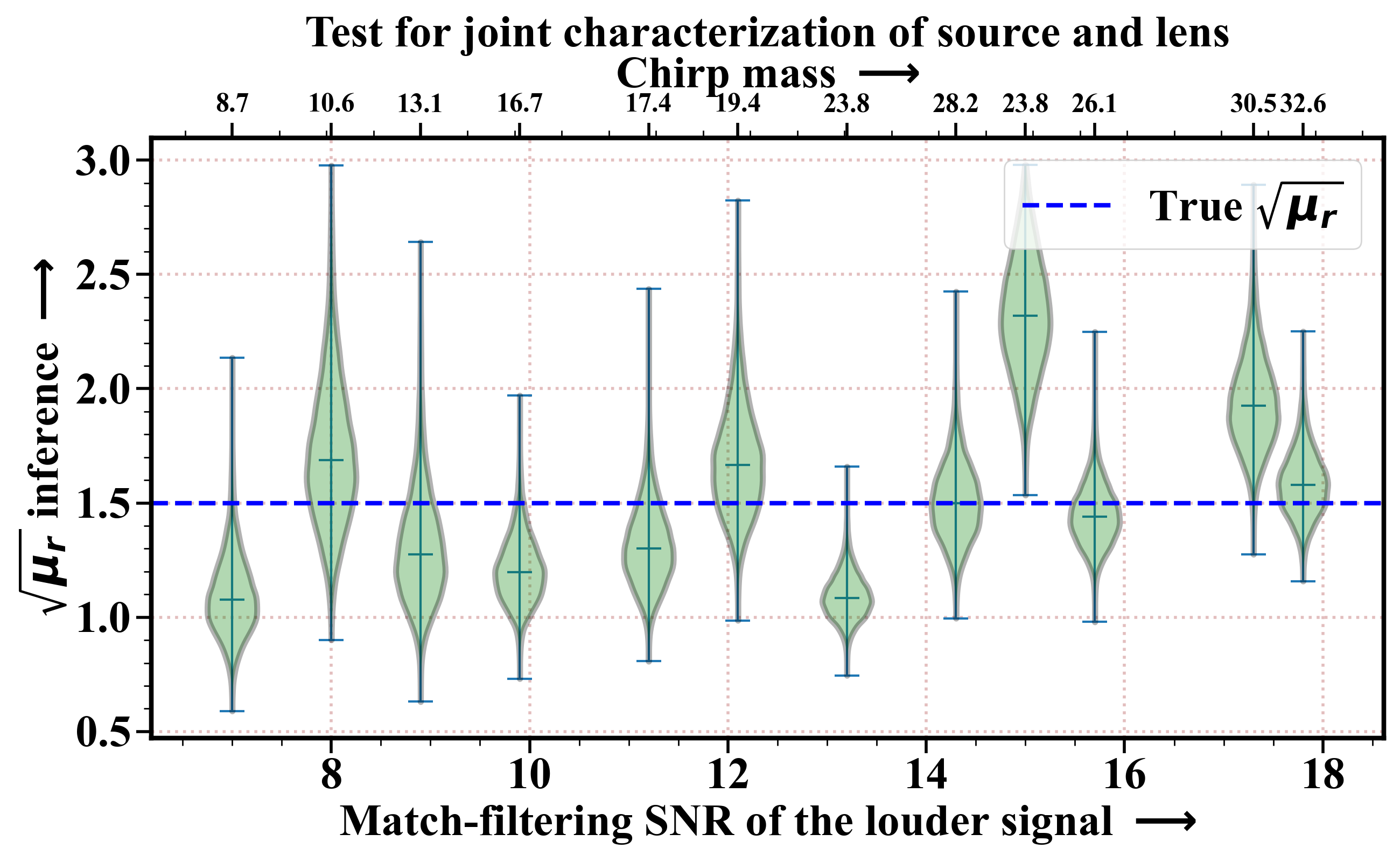}
    \caption{The figure shows the relative magnification inference for a range of chirp mass (as well as the match-filtered SNR of the louder event). The posterior distribution of the parameter is narrower around the injection value as the match-filtered SNR of the brighter source is increased.}
    \label{fig:mag_infer}
\end{figure}

Model for strong lensing: $h^l_1 (\sqrt{\mu_1}, d_l) = \sqrt{\mu _1} h_1^{ul}(d_l) $ for the higher SNR signal and $h^l_2  (\sqrt{\mu_2}, d_l)= \sqrt{\mu _2}h_2^{ul}(d_l) $ for the lower SNR signal and $d_l$, $\sqrt{\mu_1}$ and $\sqrt{\mu_2}$ are the parameters to estimate. The prior for luminosity distance is chosen to be uniform, in the range: $d_l \in [min(d_{l1}, d_{l2}), 5 \rm{Gpc}]$ where $d_{l1}$ and $d_{l2}$ are the estimated medians of the apparent luminosity distance. \footnote{The choice of $min(d_{l1}, d_{l2})$ as the lower limit in the prior describes that, if one of the two signals are magnified, we can safely say that the true luminosity distance is larger than one with minimum apparent luminosity distance. If we observe the fainter signal, there is no reason we are not going to observe the louder signal at some point of time.}. The magnifications prior is also uniform, in the range: $\sqrt{\mu_i} \in [1, 3]$ . Using Bayesian inference after the posterior is obtained, we employ a Markov Chain Monte Carlo sampler using the python-module EMCEE to obtain the posterior distribution of the parameters. We combine the posteriors of $\sqrt{\mu_1}$ and $\sqrt{\mu_2}$ to get the relative magnification $\sqrt{\mu_r} =\sqrt{\mu_2 / \mu_1}$. The results are shown in fig. \ref{fig:infer_prob} the blue lines show the injected values of the parameters. The application of this technique with the inclusion of the lensing Morse phase is discussed in appendix \ref{app:phasetest} and for the case of more than two lensing image candidates is discussed in appendix \ref{app:mul-img}. The choice of using a different, physically motivated prior in the parameter estimation is shown in appendix \ref{app:g}. The choice of using different prior led to almost identical posteriors as expected.

We tested our method on a range of cases with varying chirp mass of the BBH and checked the consistency that this method provides. The other parameters being fixed, the chirp mass itself decides the match filtering SNR of the event. Thus we have shown the variation of how well our parameter estimation works over a range of match-filtered SNRs of the louder event. The violin plot in fig \ref{fig:dist_infer} shows the inference of the luminosity distance with the increasing match-filtered SNR. The match-filtered SNR is a proxy to the strength of the signal, thus with increasing SNR, the variance of the distribution gets smaller around the injection value for the luminosity distance. Similarly, fig. \ref{fig:mag_infer} shows the performance of the parameter estimation on relative magnification, with variation in match-filtered SNR. The match-filtered SNR increase again dictates a better inference of the injection value of the $\sqrt{\mu_r}$. These trial runs test the robustness of the whole framework, before we start trial runs on the available LVK data looking for traces of lensing signature in the strong lensing regime using \texttt{GLANCE}. 

\begin{figure}
    \centering
    \includegraphics[width=0.45\textwidth]{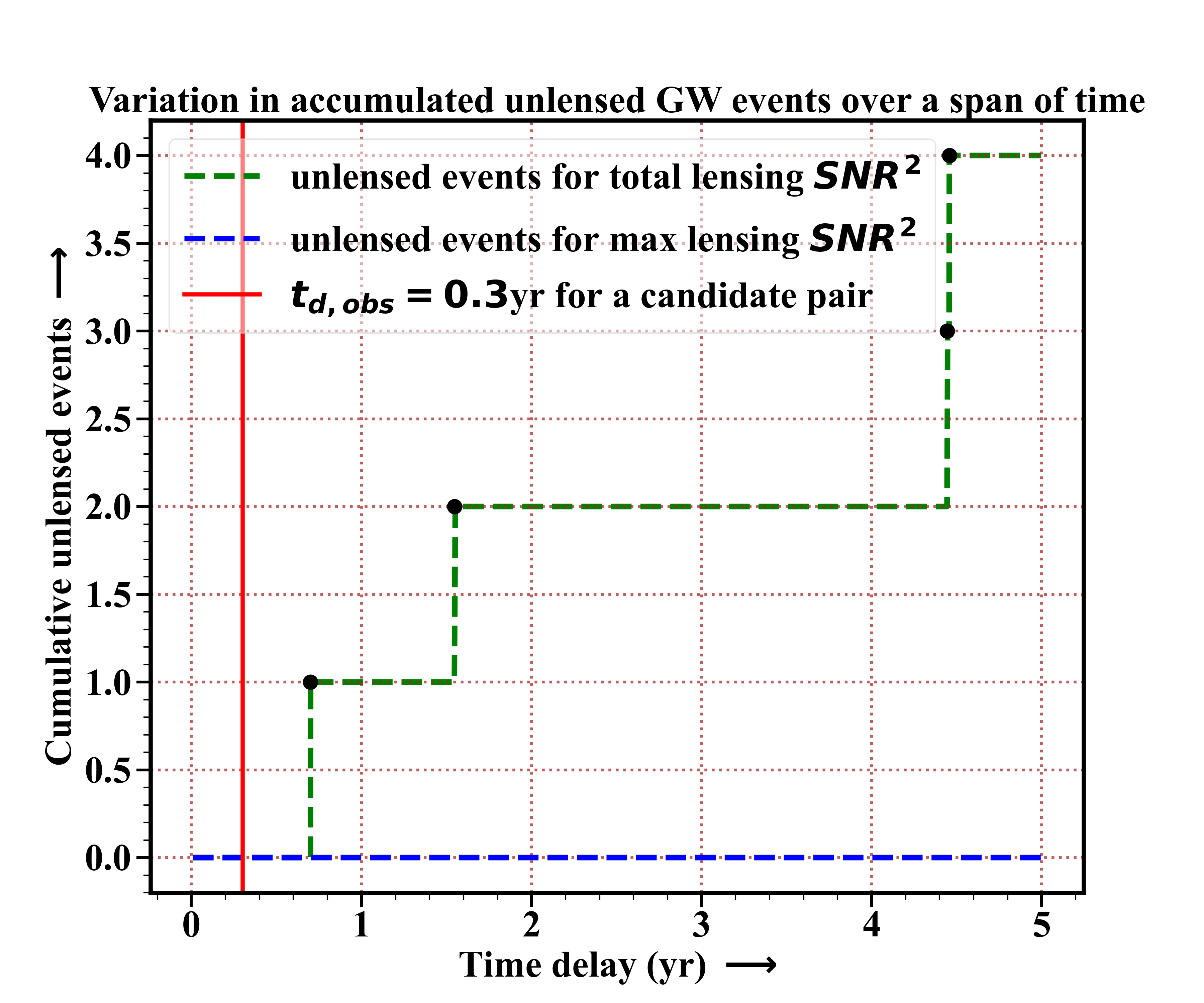}
    \caption{The figure shows the number of events accumulated over time in a certain chirp mass, spins range with overlapping sky localization. The ranges are drawn from the bounds provided by significant total lensing SNR (green) and significant maximum lensing SNR (blue). For any unlensed signal, having source parameters close to a lensed pair candidate, \texttt{GLANCE} gives high lensing SNR thus creating a false alarm for strong lensing searches. For the total lensing SNR case, given a time delay of 0.3 year between two lensed signals, no unlensed events has occurred in our simulations. However, in five years, we observe four unlensed events with similar source properties and sky localization to the lensed candidate. For the max lensing SNR case, there are no unlensed events sharing similar source properties and sky localization as the lensed candidate upto five years. The curves provide us the typical number of false lensing alarms using \texttt{GLANCE}.}
    \label{fig:FAR}
\end{figure}

\subsection{Strong lensing false alarms}

One of the crucial aspects of lensing detection is the analysis of the lensing false alarms. False alarm in strong lensing occurs when two events have very similar intrinsic source parameters and their sky localizations are overlapping. For such a case, it is hard to understand whether such two signals are truly lensed counterparts of each other or they are two unlensed signals with similar source properties and their sky localizations are overlapping just by chance. 

To demonstrate how false unlensed signal affects the strong lensing detection confirmation, we have used a PYTHON-module GWSIM \citep{Karathanasis_2023}. Given a cosmology model, BBH merger rate and black hole mass and spin distribution models, it can generate events that are detectable using current LIGO-VIRGO detectors. We have used O4 noise to obtain the number of events above individual detector match-filtering SNR $>6$. Let us consider two lensing candidates $(\rho > \rho_{th})$ are separated by a time delay of $t_d =0.3$ year. \texttt{GLANCE} predicts that the lensing SNR is significantly high over the chirp mass  $M_c \in [24,32] M_{\odot}$ and $s_{1z} \in [0.25, 0.95] $ and $ s_{2z} \in [0.25, 0.95]$, so we may get confused if the parameters lie in this range. These ranges are taken from the fig. \ref{fig18} for the total SNR case. Using the mock GW events, generated by GWSIM, over a time span of upto 5 years, we have calculated the number of events that follow the aforesaid bounds on the chirp mass and spins (provided by \texttt{GLANCE}) with overlapping sky-localizations. For two events with overlapping sky localization, we impose the condition that there is an overlap in both their RA and DEC coordinate distributions to at least 2-$\sigma$ level. We have shown the cumulative number of observed unlensed events as a function of time span between two events in fig. \ref{fig:FAR}. The plot shows that for a time delay of $0.3$ year between two lensing candidates, no detectable unlensed GW event fall in those chirp mass and spin bins with sky localization overlapping. However, we can observe that in a span of five years, there are four unlensed events sharing the candidate sky localization and intrinsic source properties. These events can trigger in false lensing alarms. However, with the choice of maximum lensing SNR (as mentioned in details in appendix \ref{app:choicesnr}), the parameter range shrinks significantly. If we observe the fig. \ref{fig:appci} and fig. \ref{fig:appcii}, \texttt{GLANCE} predicts that the maximum lensing SNR is significantly high over the chirp mass  $M_c \in [24, 27.5] M_{\odot}$ and $s_{1z} \in [0.25, 0.45] $ and $ s_{2z} \in [0.25, 0.45]$. If we keep the sky-localization overlap criteria same as before, then we observe that no unlensed event matches the the masses, spins, sky-localization of the lensed pair events. We can understand that if the delay time between two lensed GW signals is very long, the unlensed signal count (having overlap in source intrinsic and extrinsic parameters) within that time-span goes up. However, with \texttt{GLANCE}, we can get at around one unlensed per year (for total SNR-based technique) to almost no false lensing alarms (for max SNR-based technique (see appendix \ref{app:choicesnr})). To have an overview, we obtained 1322 unlensed events observed using H1-L1-V1 using O4 noise curves with individual detector SNR of six. Out of them, 61 events had masses and spins within the specified range for the total lensing SNR case. Out of them, only 4 had 2-$\sigma$ sky-overlap with the lensing candidate. However, for the max lensing SNR case, no event matched the source mass and spin criteria. So, the fraction of false alarms to the total unlensed events for total lensing SNR is $1/1322 = 0.3\%$ and for max lensing SNR it is $ \leq 1/1322 = 0.075 \%$ for five years of observation. Thus, the very small count of false alarms, proves this technique very effective in finding strongly lensed GW signals. Parameters used in the BBH merger-rate, the BBH population models and relevant parameters used in the GWSIM simulation for the FAR calculation is specified in appendix \ref{app:h}.

To summarize, we applied a cross-correlation technique to detect lensed gravitational waves. We claim that to search for lensed GW signals, a blind search on the LVK time-series data (using at least two non-coaligned) GW detectors has to be performed. Cross-correlation provides us with an insight into the degree upto which two noisy reconstruction of a polarization are similar. We have calculated the lensing SNR ($\rho$) given the signal cross-correlation and noise cross-correlation to qualify an event as lensed. If lensing SNR is greater than the threshold i.e. $\rho^2 > \rho^2_{th}$, we called it a lensing candidate. For such candidate events, we have used joint parameter estimation techniques to extract relevant source parameters degenerate with the lensing signatures on the GW waveform.

\section{Conclusion}
We have developed a new technique called \texttt{GLANCE} to find lensed gravitational wave signals. It relies on the fact that two data pieces contain GW signals with very similar characteristics, such that their overlap is a non-zero quantity. Cross-correlating noisy signal containing one polarization with un-correlated noise picks up similar signals while suppressing noise. We estimate the lensing SNR of the cross-correlation to qualify an event as a lensed candidate if the lensing SNR is above some threshold. Estimation of GW source parameters and lens parameters are performed on the candidates' events. It has been found that for above threshold ($>\rho_{th}$) SNR signals, the parameter estimation can extract the source and lens properties, and simulation results agree well with the injected values for those cases. 

To mention a few important aspects of this technique, firstly, we emphasize that \texttt{GLANCE} can work well even in the sub-threshold regime. We obtained that when the match-filtered SNR of the event is low, we still can explore the degeneracies in the parameter space but with some compromise in the precision of the results. Secondly, this technique can combine data from multiple detectors and mitigate the uncorrelated detector noise. Thus even if a detector misses to observe a sub-threshold event, we can use other detectors to get the lensing search continued. And finally, but not the least important, this technique does not depend on any particular model of lensing. It can search for any unmodelled lensed GW signal through cross-correlation between data pieces. Thus \texttt{GLANCE} can join the ongoing searches in the sub-threshold level together with the currently existing pipelines \citep{PhysRevD.102.084031, PhysRevD.107.123014}. 

One of the key aspects is to understand the false alarm rate for the detection of the lensed events. False alarms can arise when two signals even though different in source parameters, produces a high lensing SNR value. They can also arise when different events of similar source properties lie close in the sky-position. Since the GW source localization is currently not that good, false alarms can become a menace to deal with \citep{Caliskan:2022wbh}. However, The sky position of the GW source will improve drastically after LIGO-Aundha(India) becomes online \citep{shukla2023i}. With more and more detectors online like Cosmic Explorer \citep{galaxies10040090}, Einstein Telescope \citep{Punturo:2010zz} or LISA \citep{Caliskan:2022hbu}, this GW localization challenge will gradually fade away \citep{Pankow_2018}. However, within our accuracy of source localization, there can be actually two BBH-merger events with the similar source parameters, giving rise to a false lensing alarm. To resolve this, population studies of black holes have to be performed. We will explore this in details in a future work, that is currently under preparation \citep{Chakraborty:2024:newb}. 

In summary, this new technique \texttt{GLANCE} is capable of detecting lensed GW events from the noisy data for both sub-threshold and well-detected events. The application of this technique to the GW data makes it possible to detect lensed event candidates with high confidence and characterize the lens and source properties. In reality, we don't have any idea whether a signal is strongly lensed or microlensed until its source and lens parameters are completely understood. Thus in the future, this work will be extended to the micro-lensing searches using \texttt{GLANCE} \citep{Chakraborty:2024:new} to explore a a higher dimension of source parameter space degenerate the lens-imposed ones. \footnote{It will consider at least the important parameters like component masses, spins, inclination angle, and luminosity distance along with the frequency-dependent amplification caused by the lens.}. With a moderate number of GW events detectable from current-generation and next-generation GW detectors, with better noise curves, the application of this technique will push the frontier one step towards a robust detection of lensed GW signals. 

\section*{Acknowledgements}
The authors are thankful to Mick Wright for reviewing the manuscript during the LSC Publications and Presentations procedure and providing useful comments.  
We thank our colleagues for some thought-provoking insights and ideas regarding this project. This work is part of the \texttt{⟨data|theory⟩ Universe-Lab}, supported by TIFR and the Department of Atomic Energy, Government of India. The authors express gratitude to the computer cluster of \texttt{⟨data|theory⟩ Universe-Lab} for computing resources used in this analysis. We thank the LIGO-Virgo-KAGRA Scientific Collaboration for providing noise curves. LIGO, funded by the U.S. National Science Foundation (NSF), and Virgo, supported by the French CNRS, Italian INFN, and Dutch Nikhef, along with contributions from Polish and Hungarian institutes. This collaborative effort is backed by the NSF’s LIGO Laboratory, a major facility fully funded by the National Science Foundation. 

The research leverages data and software from the Gravitational Wave Open Science Center, a service provided by LIGO Laboratory, the LIGO Scientific Collaboration, Virgo Collaboration, and KAGRA. Advanced LIGO's construction and operation receive support from STFC of the UK, Max-Planck Society (MPS), and the State of Niedersachsen/Germany, with additional backing from the Australian Research Council. Virgo, affiliated with the European Gravitational Observatory (EGO), secures funding through contributions from various European institutions. Meanwhile, KAGRA's construction and operation are funded by MEXT, JSPS, NRF, MSIT, AS, and MoST. This material is based upon work supported by NSF’s LIGO Laboratory which is a major facility fully funded by the National Science Foundation.

We acknowledge the use of the following python packages in this work: NUMPY \citep{harris2020array}, SCIPY \citep{2020SciPy-NMeth}, MATPLOTLIB \citep{Hunter:2007}, GWSIM \citep{Karathanasis_2023}, ASTROPY\citep{2022ApJ...935..167A}, PYCBC \citep{alex_nitz_2024_10473621}, GWPY \citep{gwpy}, LALSUITE \citep{lalsuite}, EMCEE \citep{Foreman_Mackey_2013} and CORNER \citep{corner}, BILBY \citep{Ashton_2019}. 
\section*{Data Availability}
The data files associated with this work will be available on the \href{https://github.com/Data-Theory-Universe-Lab}{\texttt{⟨data|theory⟩ Universe-Lab}} Github page. Its usage in a research work must be done with proper consent from the authors.



\bibliographystyle{mnras}
\bibliography{mnras_template_v2} 




\appendix

\section{Basics of Strong Lensing and Micro-lensing}\label{app2}

In the presence of matter, the propagation of GW depends on the gravitational potential of intervening objects. The equation describing the frequency domain GW amplitude at a given point in space is given by \citep{1992grle.book.....S},
\begin{eqnarray}
  \left(\nabla^2 + \frac{\omega^2}{c^2} \right) \tilde{\phi} (\vec{r}, \omega) = \frac{4\omega^2 U(\vec{r})}{c^4} \tilde{\phi}(\vec{r}, \omega),
\end{eqnarray}
where $\omega$ is the frequency ($=2\pi f$) of the gravitational wave, $U(\vec{r})$ is the time-independent gravitational potential of the lensing massive object \footnote{This equation assumes that the lensing potential does not vary with time.} and $h_{\mu \nu } (\omega, \vec{r}) = e_{\mu \nu} \tilde{\phi}(\omega, \vec{r})$ such that $ \tilde{\phi}(\omega, \vec{r})$ is the frequency domain amplitude of GW and $e_{\mu \nu}$ is the polarization information.

In the presence of a lensing object, GW geodesics are modified, which leads to a change in the trajectory of the signal and causes a time delay in the arrival of the different lensed signals. These effects altogether lead to the diffraction of gravitational waves quite similar to EM-wave optics \citep{Meena_2019, PhysRevD.108.043527}. 

\begin{figure}
    \centering
    \includegraphics[width=0.48\textwidth]{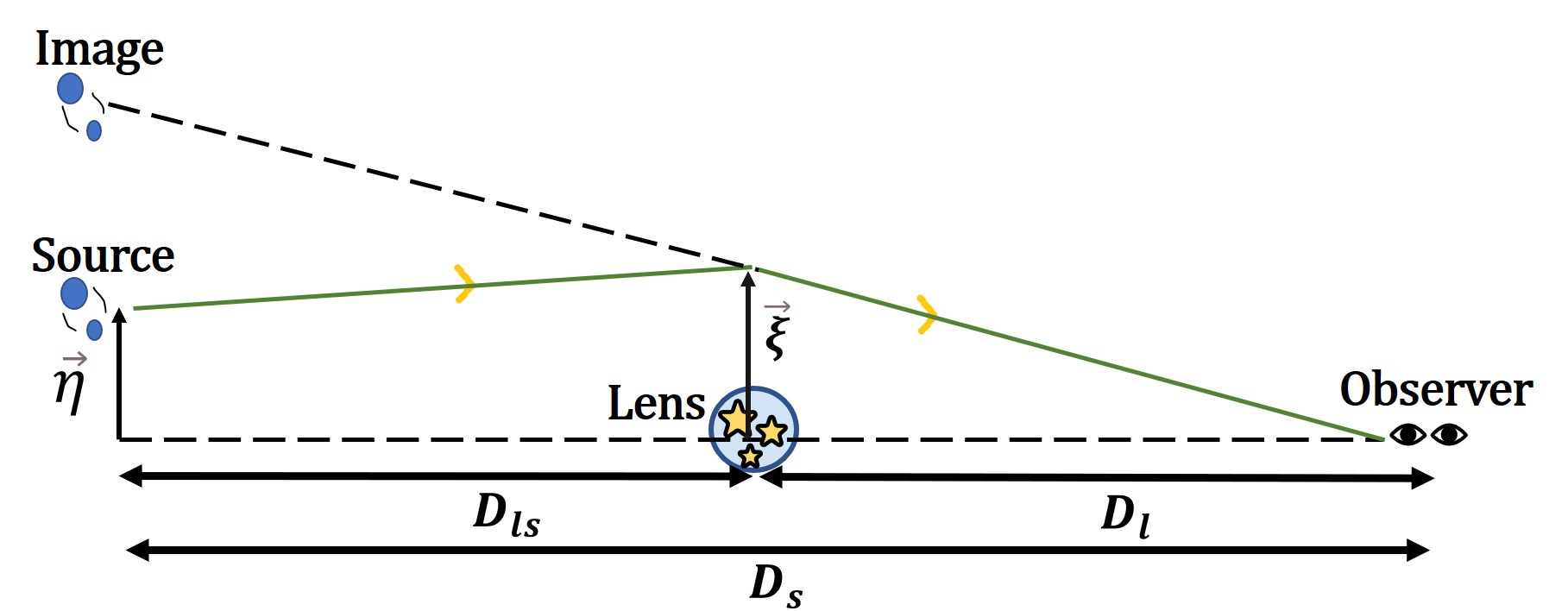}
    \caption{The figure shows the ray diagram schematic for lensing of gravitational waves by a thin lens approximated object. A thin lens approximates that the deflection of the ray caused by the lensing objects happens only at the lens plane.}
    \label{fig4}
\end{figure}

In fig. \ref{fig4} we have a schematic diagram of lensing by a thin lens. In thin lens approximation we assume that the deflection of the GW caused by the massive lensing object happens in the lens plane only. Here $D_l$ is the distance between the observer and the lens, $D_s$ is the distance between the observer and the source and $D_{ls}$ is the distance between the lens and the source. $\vec{\eta}$ is the position of the source in the source plane from the intersection point between the straight line connecting the lens-center to the observer and the source-plane. $\vec{\xi}$ is the position connecting the lens center to the GW trajectory incidence point in the lens-plane, we have defined two dimensionless vectors $\vec{x}$ and $\vec{y}$ as following:
\begin{eqnarray}
    \vec{x} = \frac{\vec{\xi} } {\xi_0} \quad \text{and} \quad \vec{y} = \frac{ D_{l} \vec{\eta}}{\xi_0 D_s},
\end{eqnarray}
where $\xi_0$ is a characteristic length of the system, called the Einstein radius. A thin lens is a fair approximation that is used when the dimension of the lens is much smaller compared to the total traversed distance by the GW. Depending on the size of the lensing object as compared to the wavelength of GW, we can categorize lensing scenarios into strong lensing or micro-lensing. The frequency-domain amplification factor $F(f) = \frac{\tilde{\phi}^{\rm lensed}(f, \vec{r})}{\tilde{\phi}^{\rm unlensed} (f, \vec{r})}$, where $\tilde{\phi}$ is mentioned above, is dependent on the relative time delay between the arrival of GWs through different trajectories. The time delay incorporates geometric time delay, caused by the time difference between the GW paths, and Shapiro time delay, caused by the general relativistic time lag correction between different GW paths in curved spacetime. The total time delay is given by \citep{1992grle.book.....S, Takahashi_2003, universe7120502, Bulashenko_2022, Grespan:2023cpa}
\begin{equation}
    t_d(\vec{x}, \vec{y})=\dfrac{D_s \xi_0^2}{c D_l D_{ls}}\left(1+z_l\right)\left[\frac{1}{2}|\vec{x}-\vec{y}|^2-\psi(\vec{x})+\Phi_m(\vec{y})\right] \quad,
    \label{eq1}
\end{equation}
where $z_l$ is the redshift of the lens. The first term gives the geometric time-delay and the second term gives with the Shapiro time-delay where $\psi(\vec{x})$ gives the gravitational potential of the lens in the 2D lens plane. The third term with $\Phi_m(\vec{y})$ is required to set the minimum $t_d$ value to zero. For any wave passing through an aperture, the amplification factor is obtained from the Kirchhoff integral. Using the time-delay expression mentioned above, we can find the amplification factor as, 
\begin{equation}
    F(f)=\frac{D_s \xi_0^2}{c D_l D_{ls}} \frac{f}{i} \int d^2 \vec{x} \exp \left[2 \pi i f t_d(\vec{x}, \vec{y})\right],
\end{equation}
Accounting for the cosmological expansion, we can replace all frequencies $f$'s by $f(1+z_l)$ where $z_l$ is the redshift of the lens. For a radially symmetric lens potential, the expression can be written as 
\begin{equation}
    F(w)=-i w e^{i w y^2 / 2} \int_0^{\infty} dx \left[x J_0(w x y) e^{ \left[i w\left(\frac{1}{2} x^2-\psi(x)+\Phi_m(y)\right)\right]} \right],
\end{equation} 
here $J_0$ is the spherical Bessel function of zeroth order, and $w$ is the dimensionless frequency defined as, $w= 8 \pi G M_{lz} f/c^3$ where $M_{lz} = M_l (1+z_l)$ is the redshifted mass of the lens, $f$ be the frequency of the GW, $G$ be the gravitational constant and c is the speed of light in vacuum.

Analytical solutions can be obtained when the lens potential is spherically symmetric. For a commonly used galaxy/cluster density model called singular isothermal sphere profile, the amplification factor is given by \citep{PhysRevD.108.043527},

\begin{equation}
    F(w) = e^{\frac{iwy^2}{2}} \sum_{n=0} ^{\infty} \frac{\Gamma(1+\frac{n}{2})}{n!} (2we^{\frac{i3\pi}{2}})^{\frac{n}{2}}{_1\tilde{F}_1} (1+\frac{n}{2}, 1, -\frac{iwy^2}{2}),  
\end{equation}
where $\Gamma $ denotes the gamma function and $_1\tilde{F}_1$ is the confluent hyper-geometric function. 

\begin{figure}
  \centering
  \includegraphics[width=0.48\textwidth]{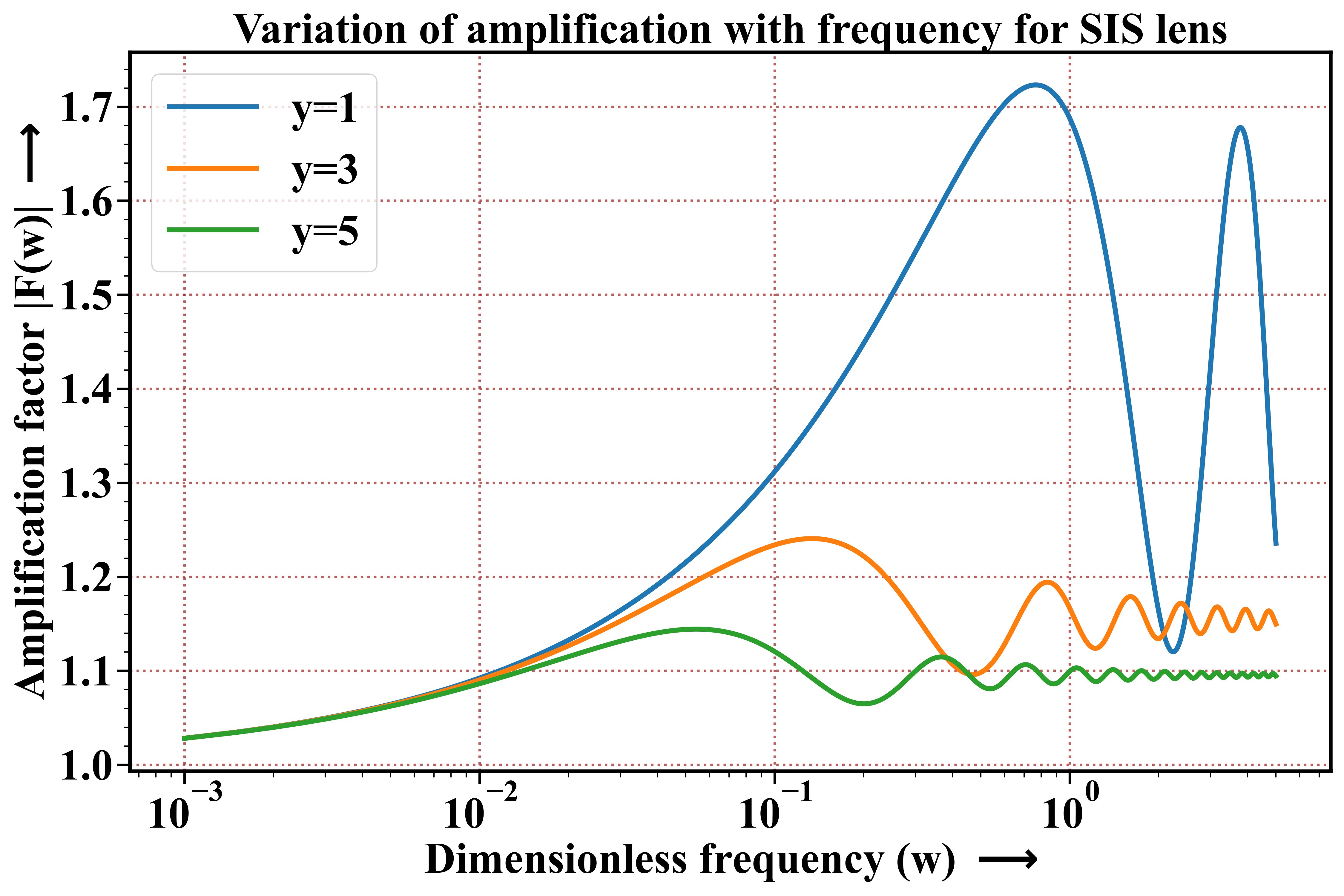}
  \caption{The figure shows the variation of amplification factor for gravitational lensing with respect to dimensionless frequency for a singular isothermal sphere (SIS) lens with a range of impact parameters (y's). In the condition $w \lesssim 1$ (wave optics regime, called microlensing), the amplitude of the amplification factor oscillates rapidly. However, as we approach the limit with $w \gg 1$ (geometric optics regime, called strong lensing), it converges to a single value. It can be shown that a similar trend is followed by the phase of the amplification factor.}
  \label{fig4b}
\end{figure}

We have plotted the amplitude of the complex amplification factor ($F(w)$) vs the dimensionless frequency ($w$) for several impact parameters $y$'s in fig. \ref{fig4b}. We can observe that, as the frequency of the gravitational wave is increased, at $w \gg 1$, the wavelength of the GW gets much shorter than the Schwarzschild radius of the lens and we are in the geometric optics regime. In this regime, $|F(w)|$ tends to converge at a single value, showing tiny to almost no variations in $w$, leading to no significant frequency-dependent amplitude modulation of the GW. This is the strong lensing regime of GW. However, in the range $w \lesssim 1$, the wavelength is of comparable size or larger than the Schwarzschild radius of the lens and we are in the wave optics regime. Here, $|F(w)|$ shows a strong dependency with $w$, both its amplitude is highly oscillatory with respect to variations in $w$ \citep{Matsunaga:2006uc, PhysRevLett.122.041103, PhysRevD.90.062003}, leading to the frequency-dependent modifications of the GW. This is the microlensing regime of GW. It can be shown similarly, that the phase of the amplification factor ($\theta_F (w) = - i ln \left(\frac{F(w)}{\lvert F(w) \rvert} \right)$) follows a similar trend to the amplitude of $F(w)$, showing frequency-independent phase modulation for $w \gg 1$ and strong frequency-dependent oscillatory nature for $w \lesssim 1$ \citep{Takahashi_2003, 
PhysRevD.108.103532}. It can be shown that, for any generic lens system, the amplification factor in the strong lensing regime becomes, 

$$
F(f)=\sum_j\left|\mu_j\right|^{1 / 2} \exp \left[2 \pi i f t_{d, j}-i \pi n_j\right],
$$
where, the magnification of the j-th image is $\mu_j = 1 / \operatorname{det}\left(\frac{\partial \vec{y}}{\partial \vec{x}_j}\right)$, and  $t_{d,j} = t_d (\vec{x_j}, \vec{y})$ and $n = 0, 1/2, 1$ for minimum, saddle point, maximum of the $t_d(\vec{x}, \vec{y})$ function. They are known as type-I, type-II and type-III images respectively. The time-domain lensed wave amplitude can be represented as, 

$$
\phi^{ \rm lensed} (t, \vec{r})=\sum_j\left|\mu_j\right|^{1 / 2} \phi^ {\rm unlensed}\left(t-t_{d, j}, \vec{r}\right) \exp \left[-i \pi n_j\right]
$$

Thus the amplitude and phase modulation in the strong-lensing regime is independent of frequency \citep{dai2017waveforms}. We obtain a magnified time-domain waveform with zero, $\pi/2$ or $\pi$ phase modulation for strong lensing. In this project, we assume, $\lambda \gg R^s _{\rm lens}$ and work in the strong lensing scenario. However, since GW-frequency evolves over time, a lensed GW signal may show a transition from the wave optics to the geometric optics regime depending on the mass of the lens and the wavelength of the GW.

\section{Signal reconstruction from data}\label{app:signal_reconstruction}
\begin{figure*}
    \centering
    \includegraphics[width=0.95\textwidth]{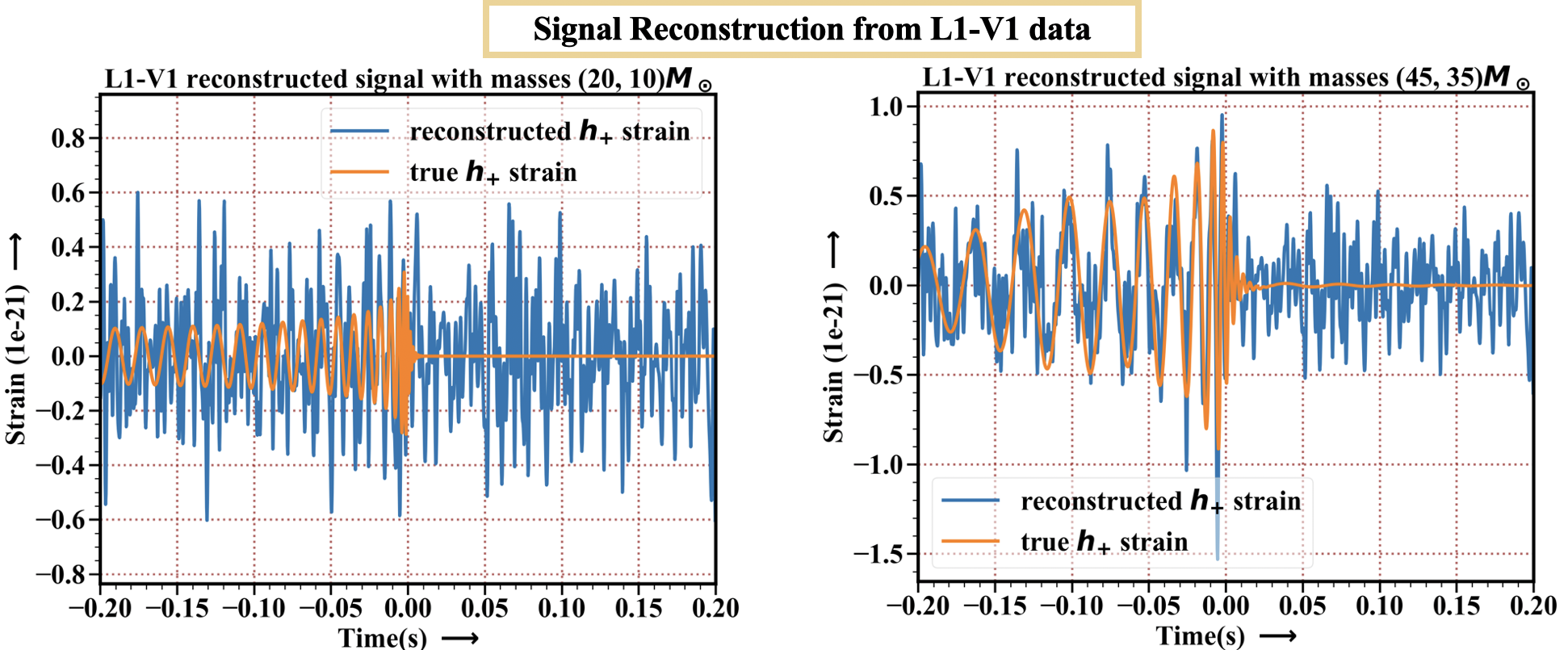}
    \caption{The figure shows the plus polarization signal reconstruction from Livingston-L1 and Virgo-V1 data. We can observe the with increase in the component masses of the BBH system, the plus polarization signal is better recovered from the data.} 
    \label{fig:sig_recon}
\end{figure*}

As mentioned in eq. \ref{data}, the data consists of the signal and the noise, whereas the signal constitutes from each of the GW polarization times the antenna response function of a detector towards the source position in the sky.

Before cross-correlation, we process the data first by passing through a band pass filter, which only allows a frequency band 30Hz-512Hz. Applying a 30Hz cutoff, high pass filter is necessary to remove some of the dominant modes in terrestrial seismic noise, shot noise, thermal noise etc. The low pass filter with cutoff at 512Hz removes excess jitter and smoothes the data out. \footnote{For low mass coalescence events ($M_{\rm total} < 10 M_{\odot}$), a cutoff at 1024Hz can also be considered. However, if higher order modes are contributing significantly in the strain at high frequencies, we may need to incorporate even higher cutoff frequency.}

Note that, the signal arrives at slightly different time at different detectors. The time delay depends upon the source position in the sky. We pick a possible sky location of the source (for events we can rely on GRACEDB i.e. Gravitational-Wave Candidate Event Database), otherwise through some intelligent guess), and estimate the time delay between detectors. Accounting that as a time delay, we pick a start GPS-time shifted by the same amount as the time delay in the detector with respect to a reference detector. We then chop a 16s long strain data for each of the detectors. This makes the signal, if present, exactly at equal time after the start time of the chopped data for each detectors. 
To obtain the best-reconstructed signal containing only one polarization
we formulate a matrix equation for these data chunks\footnote{$d_1$ and $d_2$ are slightly time shifted due to the reason mentioned before.}, 

\begin{eqnarray}
\begin{pmatrix}
d_1\\
d_2
\end{pmatrix} =
\begin{pmatrix}
F_{+ 1} & F_{\times 1}\\
F_{+ 2} & F_{\times 2}
\end{pmatrix} 
\begin{pmatrix}
h_{+}\\
h_{\times}
\end{pmatrix}
+
\begin{pmatrix}
n_1\\
n_2
\end{pmatrix}
\end{eqnarray}

To solve for the two polarization signals, we invert the antenna pattern matrix and multiply it with the column vector at the left. Therefore, we obtain,

\begin{eqnarray}
\centering
\begin{pmatrix}
F_{+ 1} & F_{\times 1}\\
F_{+ 2} & F_{\times 2}
\end{pmatrix} ^{-1} 
\begin{pmatrix}
d_1\\
d_2
\end{pmatrix} =
\begin{pmatrix}
d^{+}_{12}\\
d^{\times}_{12}
\end{pmatrix}
=
\begin{pmatrix}
h^{+}_{12}\\
h^{\times}_{12}
\end{pmatrix} +
\begin{pmatrix}
n^{+}_{12}\\
n^{\times}_{12}
\end{pmatrix},
\end{eqnarray}

where, 
$$
\begin{pmatrix}
n^{+}_{12}\\
n^{\times}_{12}
\end{pmatrix}
= \begin{pmatrix}
F_{+ 1} & F_{\times 1}\\
F_{+ 2} & F_{\times 2}
\end{pmatrix} ^{-1} 
\begin{pmatrix}
n_1\\
n_2
\end{pmatrix}.
$$
We have renamed $h_{+}$ and $h_{\times}$ to $h^{+}_{12}$ and $h^{\times}_{12}$ respectively, to denote that they are constructed using detectors 1 and 2. 
In eq. \ref{eq:reconstructed_data}, we used this reconstructed noisy signal $d^{+}_{12}$ where we also renamed `12' to `x' as a naming convention for ease.
We use these best reconstructed signals for cross-correlation analysis to search for strong lensing. In fig. \ref{fig:sig_recon}, we have demonstrated the plus polarization signal reconstruction using LIGO-Livingston (L1) and VIRGO (V1) data for two different choices of BBH component masses. All these sources are placed at 1Gpc and are not magnified at all. As the signal strength goes stronger (increased match-filtering SNR), the sky localization errors get smaller and the polarization signals are better reconstructed. Also this signal reconstruction technique depends of the known values of antenna pattern, as a function of GPS time, polarization angle, RA and DEC coordinates. So if the GW event is detected through at least three detectors (with the chirp mass and aligned spins falling in the range discussed in the false alarm rate curve in section \ref{section6} ) mean sky localization of an event is around 85 sq. degrees at $90\%$ CI. Pinpointing the GW source position on the skymap will further improve with the upgradation LIGO detectors in O5 as well as with the inclusion of next-generation detectors like Cosmic Explorer or Einstein Telescope. For such events, we can pick a point within the 90\% CI contour of the sky localization and can find both the polarization signals, with antenna functions calculated at that point. \footnote{It matters very less if the actual position of the GW event on the sky is not at those medians since the antenna pattern is hardly varying within such small sky patches} . For broad sky-localization events, we have to find the antenna function at many different points on the RA-DEC coordinate grid well distributed throughout the $90\%$ CI skypatch. The source sky location is then decided by the position at which the cross-correlation on the reconstructed signals is maximum.

\section{Off-phase cross-correlation of two signals, noise cross-correlation with signal, phase shifted signal cross-correlation: Three essential tests for the \texttt{GLANCE} framework}\label{app:phasetest}

\begin{figure}
    \centering
    \includegraphics[width=0.48\textwidth]{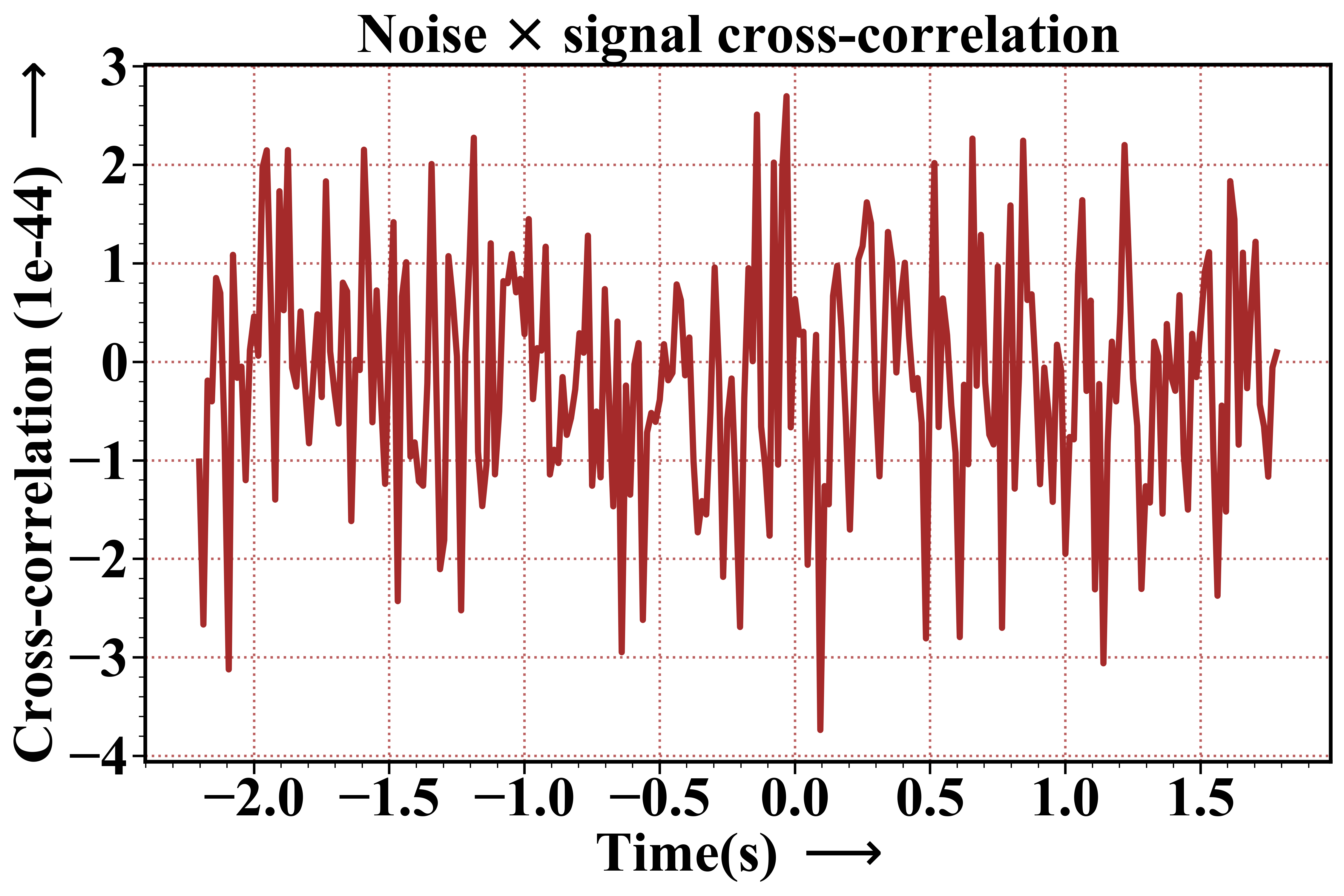}
    \caption{The figure shows the cross-correlation between noise and signal shows no evidence of a trend similar to signal cross-correlation. Here the cross-correlation timescale is chosen to be $\tau_{cc} =0.0625s$.}
    \label{fig:appdi}
\end{figure}

We have observed the cross-correlation of noise with itself (at two different instants) in fig. \ref{fig6} and the cross-correlation of a signal with another signal in fig. \ref{fig6a}. However, one may wish to explore how the signal cross-correlated with noise appears. Keeping this in mind, we have shown the same in fig. \ref{fig:appdi}.

\begin{figure}
    \centering
    \includegraphics[width=0.48\textwidth]{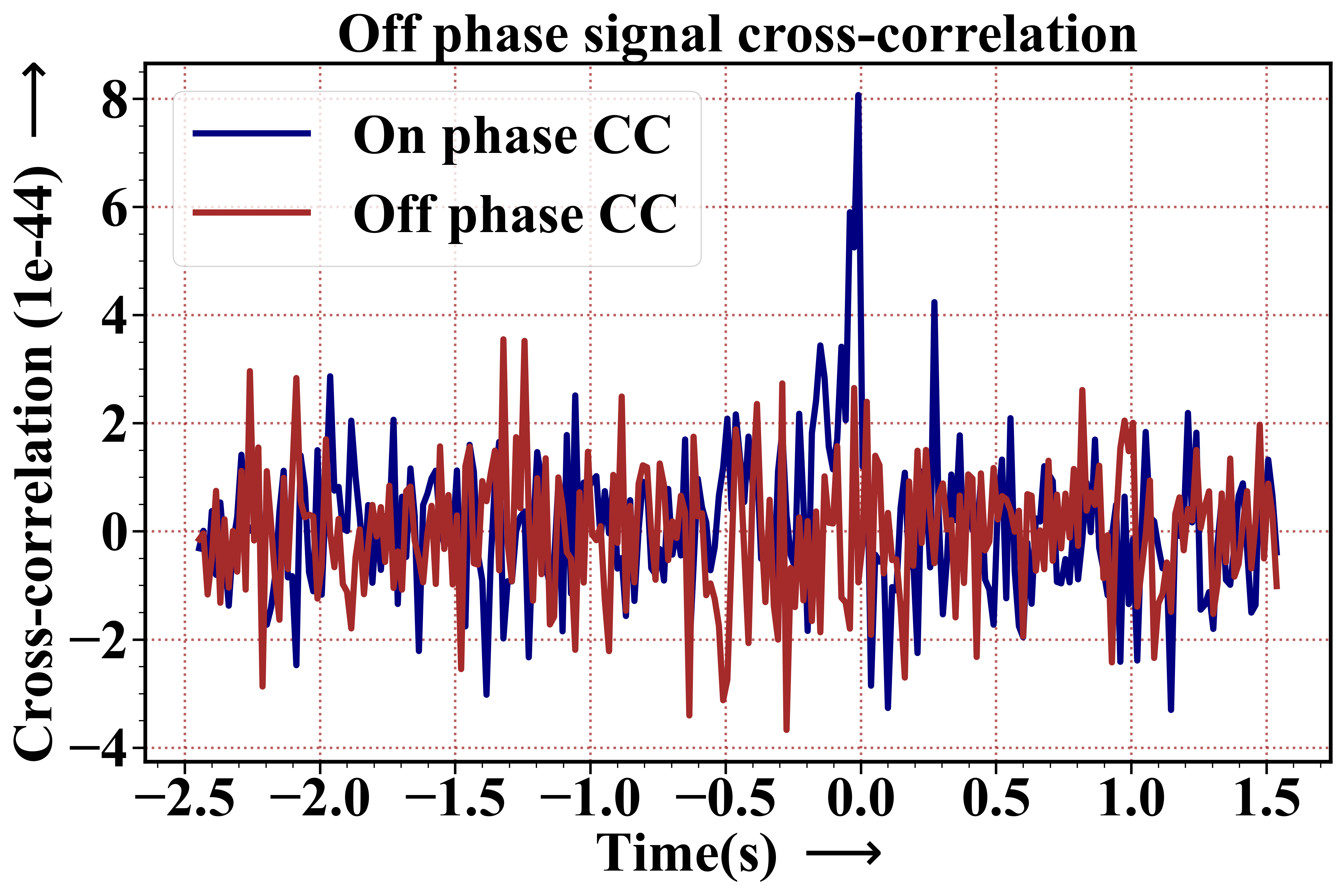}
    \caption{The figure shows the cross-correlation between two signals, but not on phase with $\tau_{cc}= 0.0625s$ vs cross-correlation of two signals on phase. It is evident that non-overlapping cross-correlation does not produce a nice cross-correlation signal.}
    \label{fig:appdii}
\end{figure}

\begin{figure}
    \centering
    \includegraphics[width=0.48\textwidth]{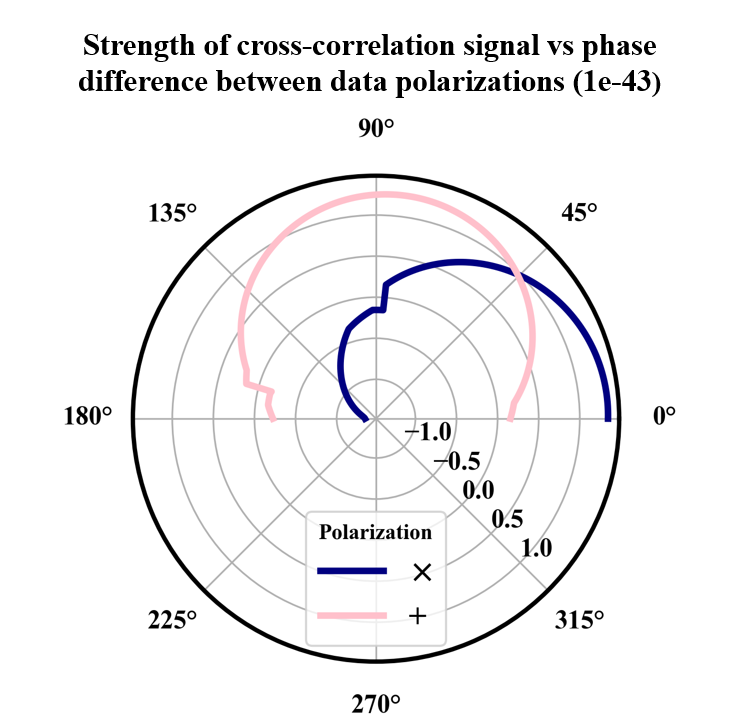}
    \caption{The figure shows the polar plot showing the peak strengths of the cross-correlation with variation in Morse phase (due to lensing) of the second GW signals. It shows the peak cross-correlation between a cross and a plus polarization (coloured pink) and the cross-correlation between a cross and another cross polarization signal (coloured navy). Starting from the phase difference of zero, the cross-correlation with the cross-polarization signal goes down monotonically becoming negative after a phase-diff of $\pi/2$. The cross-correlation with the plus polarization signal, however, first increases, becomes maximum at $\pi/2$ then decreases. In type-II images, the lensed signals have a phase difference of $\pi/2$. When the cross-correlation between two lensed cross-polarization signals is close to zero (at phase shift of $\pi/2$), the cross-correlation with the plus-polarization signal is maximum. Thus if we cross-correlate any of the polarizations of one lensed signal with both the polarizations of another lensed signal, we can retrieve the lensed signals by cross-correlation. The curve shows a sudden jump near the zero-crossings when the noise cross-correlation is larger than the signal cross-correlation.}
    \label{fig:appdiii}
\end{figure}

\begin{figure}
    \centering
    \includegraphics[width=0.48\textwidth]{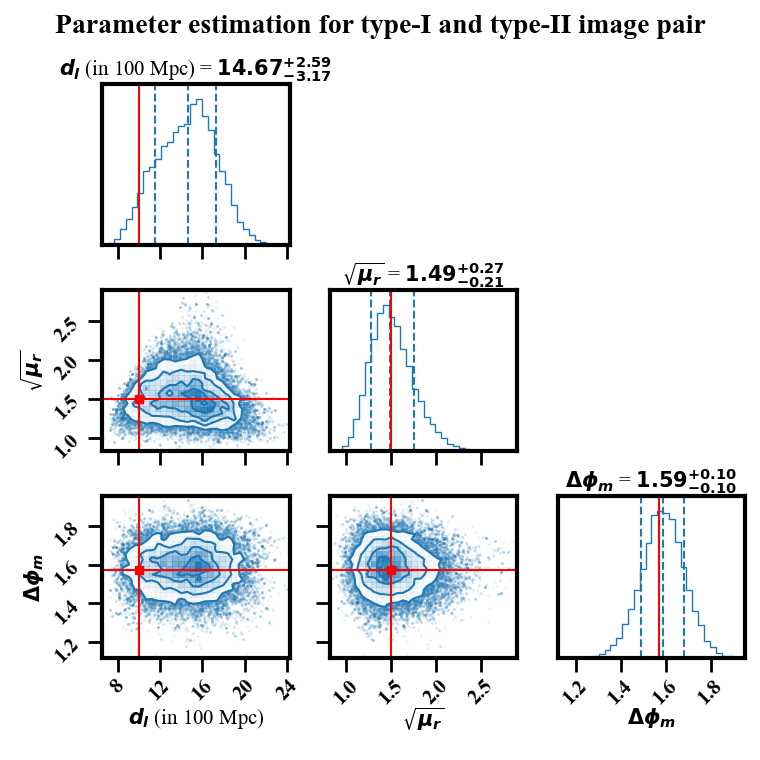}
    \caption{The figure shows the source and lensing joint parameter estimation for a type-I and type-II image pair. There is an expected Morse phase shift of  $\pi/2$ which is reflected in the Morse phase difference parameter $\Delta \phi_m$. The choices of the other source and lensing parameters are kept the same as in the case of section \ref{section6}.}
    \label{fig:morse_phase_param_estim}
\end{figure}

As it can be seen here, we cannot distinguish any pattern similar to fig. \ref{fig6a}. We are safe to say that, noise cross-correlated with the reconstructed signal does not show any cross-correlation pattern. Therefore, such cases won't pose a threat to this technique generally. 

While, writing eq. \ref{eq7}, we essentially assumed that we have already found two signals separated by a time delay of $t_d = t_{d0}$. However, as we mentioned there, in practical cases we are running a blind search for other signals, making $t_d$ a free parameter. In some cases, when the signals have some overlap but not lying exactly on top of one another, we may have a mild-cross-correlation signal. To check, whether we would receive a strong signal cross-correlation or not, we have plotted the cross-correlation of two signals slightly off phase in fig. \ref{fig:appdii}. Here the maximums of the two signals are separated by 0.244s, with signals each having a duration of about 1s. We can observe that this off-phase cross-correlation appears similar to the noise cross-correlation in fig. \ref{fig6}, whereas the on-phase cross-correlation (with maximums of the signals lying on top of each other) shows a distinguishable peak. 

Additionally, we have shown that if there is a phase shift of the signals due to lensing, how that is going to affect the signal cross-correlation. Suppose that in type-II image the signal is impacted by a phase shift of $\pi/2$. If we take the cross-polarization piece from the data and cross-correlate it with both the polarizations of another reconstructed signal, we would see that now the cross-polarization signal nicely overlaps with the plus-polarization signal of the second image. In fact fig. \ref{fig:appdiii} shows strength the cross-correlation signal are quite complementary when plotted against phase-shift. In the plot we took a noisy cross-polarization signal and cross-correlated it with a noisy plus and a noisy cross polarization with different phase shifts. For type-II images, the phase shift is $\pi/2$. It is easily understood that although the cross-correlation between the two plus polarizations fall drastically at this case, the cross-correlation between the cross and the plus polarization is maximum. Here the domain for the phase-shift is chosen to be $\theta = [0, \pi]$. The cross-polarization signal strength monotonically goes down with increasing phase difference, crossing zero and reaching negative values after the phase difference of $\pi/2$. The plus polarization cross-correlation first increases with a phase difference, becomes maximum at $\pi/2$, and then goes down towards zero at $\pi$. This shows that if there is a phase shift of $\pi/2$ between the two $h_+$ signals (or $h_\times$) GW signal due to lensing, then $h_\times$ (or $h_+$) polarization signal can be used for cross-correlation. Thus as a fail-safe method, we always cross-correlate each polarization of an image signal with both of the polarizations of the other image signal. These three tests are a must to check the consistency of the whole mathematical framework behind \texttt{GLANCE}. Our tests show that the technique passes well through these tests. Thus when this technique shows distinguishable cross-correlation peak, we check the origin of this cross-correlation signal, by evaluating its SNR and running a parameter estimation with those two data-pieces.

The strength of the cross-correlation signal when two signals are not lensed i.e. have different parameters is skipped, because we already have a similar plot i.e. fig. \ref{fig18}. It shows that when the parameters of the two signals match (apart from luminosity distance), we obtain the tallest peak of the summed lensing SNR and the total lensing SNR falls off quite quickly as we change the value of the parameter.

The parameter estimation technique for a pair of type-I and type-II images needs slight modification. The lensing Morse-phase shift is equivalent to a change of GW coalescence phase by half of that amount. Therefore, by we need to explore the coalescence phase parameter together with luminosity distance and magnifications for a full-parameter estimation. If we follow a similar approach to section \ref{section6} with the same injection parameters used there, we can keep the same equations upto \ref{eq:data} with the likelihood as
\begin{equation}\label{eq:likelihood2}
\begin{split}
    -\log(L) &= \frac{(d_{1} - h^{l}_1 (\sqrt{\mu_1}, d_l, \phi_{c1}))^2}{2\sigma_{n1}^2} + \frac{\log(2 \pi \sigma_{n1}^2)}{2} \\
    &+ \frac{(d_{2} - h^{l}_2 (\sqrt{\mu_2}, d_l, \phi_{c2}))^2}{2\sigma_{n2}^2} + \frac{\log(2 \pi \sigma_{n2}^2)}{2}.
\end{split}
\end{equation}

The models now consist of additional coalescence phase parameter:
$h^{l}_1 (\sqrt{\mu_1}, d_l, \phi_{c1})= \sqrt{\mu _1} h_1^{ul}(d_l, \phi_{c1})$ and $h^{l}_2 (\sqrt{\mu_2}, d_l, \phi_{c2})= \sqrt{\mu _2} h_2^{ul}(d_l, \phi_{c2})$. The choices of the priors are uniform $d_l \in [min(d_{l1}, d_{l2}), 5 \rm{Gpc}]$ where $d_{l1}$ and $d_{l2}$ are the estimated medians of the apparent luminosity distance and magnifications $\sqrt{\mu_i} \in [1,3]$. Since the Morse phase of a lensed signal can be $0$, $\pi/2$ or $\pi$ (for type-I, II and III images respectively), the equivalent coalescence phases corresponding to them are $0$, $\pi/4$ and $\pi/2$ for the reason discussed before. Therefore, we choose the coalescence phase priors as uniform within the range $\phi_{ci} \in [0, \pi/2]$. We then reduce the samples to $\sqrt{\mu_r}$ as discussed in section \ref{section6} now accompanied by another sample reduction of Morse phase difference between the images as $\Delta \phi_m = 2 \times (\phi_{c2} - \phi_{c1})$. The parameter estimation for a pair of type-I and type-II image is shown in fig. \ref{fig:morse_phase_param_estim}. The figure shows that the Morse-phase shift, if exists between two images, can be recovered using the joint PE algorithm.

\section{Formalism for parameter estimation in case of multiple images}\label{app:mul-img}
\begin{figure}
    \centering
    \includegraphics[width=0.48\textwidth]{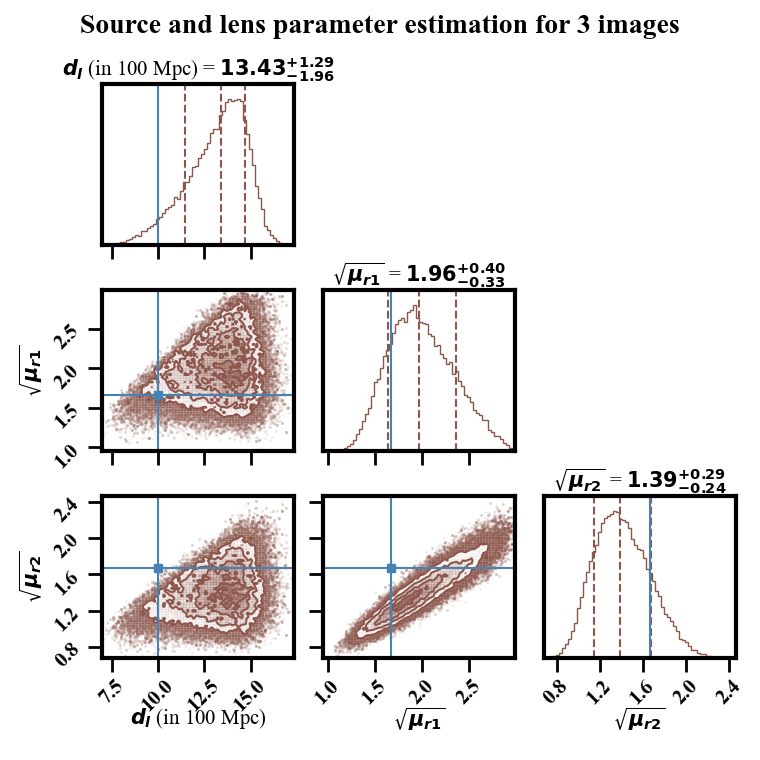}
    \caption{The figure shows the inference of relative magnifications and luminosity distance for a 3-image scenario. The source and lens parameters are again degenerate; with three image signals, we can infer up to three parameters, here $d_l$ i.e. the true luminosity distance of the source, $\sqrt{\mu_{r2}}$ and $\sqrt{\mu_{r3}}$ giving the relative magnification of the second and third images with respect to the first image.}
    \label{fig:appb}
\end{figure}
The cross-correlation technique can be applied between any two data pieces together and the lensing SNR would tell us about the significance of the event as lensed. However, application of \texttt{GLANCE} for detection of lensed events can be extended to cases with more than two images, handling one pair at a time. However, the parameter estimation with three strongly lensed images, we have three data pieces and then try to infer four parameters ($\sqrt{\mu_1}, \sqrt{\mu_2}, \sqrt{\mu_3}, d_l$) shows strong degeneracy in those 4 parameters. 

Thus we formulate a new likelihood function that takes these three data pieces as inputs and infers three parameters from it. These three parameters are: $d_l$ i.e. the true luminosity distance of the source, $\sqrt{\mu_{r2}}\equiv\sqrt{\mu_2 /\mu_1}$ and $\sqrt{\mu_{r3}}\equiv\sqrt{\mu_3 /\mu_1}$ giving the relative magnification of the second and third images with respect to the first image. The likelihood function now appears as,

\begin{align}
    -\log(L) &= \frac{(d_{1} - h^{l}_1 (\sqrt{\mu_1}, d_l))^2}{2\sigma_{n1}^2} + \frac{\log(2 \pi \sigma_{n1}^2)}{2}  \nonumber\\ & + \frac{(d_{2} - h^{l}_2(\sqrt{\mu_2}, d_l))^2}{2\sigma_{n2}^2} + \frac{\log(2 \pi \sigma_{n2}^2)}{2} \nonumber\\
    &+ \frac{(d_{3} - h^{l}_3(\sqrt{\mu_3}, d_l))^2}{2\sigma_{n3}^2} + \frac{\log(2 \pi \sigma_{n3}^2)}{2}
\end{align}
with $h^l_1(\sqrt{\mu_1}, d_l) = \sqrt{\mu_{1}} h_1^{ul}(d_l) $  and $h^l_2(\sqrt{\mu_2}, d_l) = \sqrt{\mu_{2}} h_2^{ul}(d_l) $ and $h^l_3(\sqrt{\mu_2}, d_l) = \sqrt{\mu_{3}} h_3^{ul}(d_l) $ where the notations follow from equations \ref{eq:data} and \ref{eq:likelihood}. Like before, we obtained the posteriors for these four parameters and combine the samples to create distribution for three reduced parameters: $d_l$ , $\sqrt{\mu_{r2}} =\sqrt{\mu_2 /\mu_1}$ and $\sqrt{\mu_{r3}} = \sqrt{\mu_3 /\mu_1}$. In fig. \ref{fig:appb}, we have shown the posterior distribution of those parameters. The recovery of the lensed signal source parameter ($d_l$) is possible through this association of the lens-imposed magnifications ($\sqrt{\mu_{ri}}'s$). We can further include the Morse phase as an explorable parameter to measure the relative Morse phases between signals as shown in appendix \ref{app:phasetest}.
This technique can be extended to cases with any number of images with subsequent modifications needed in the likelihood function.

\section{Different noise realizations: how that throws off the lensing SNR and the parameter estimation}\label{app:noise}

\begin{figure}
    \centering
    \includegraphics[width=0.48\textwidth]{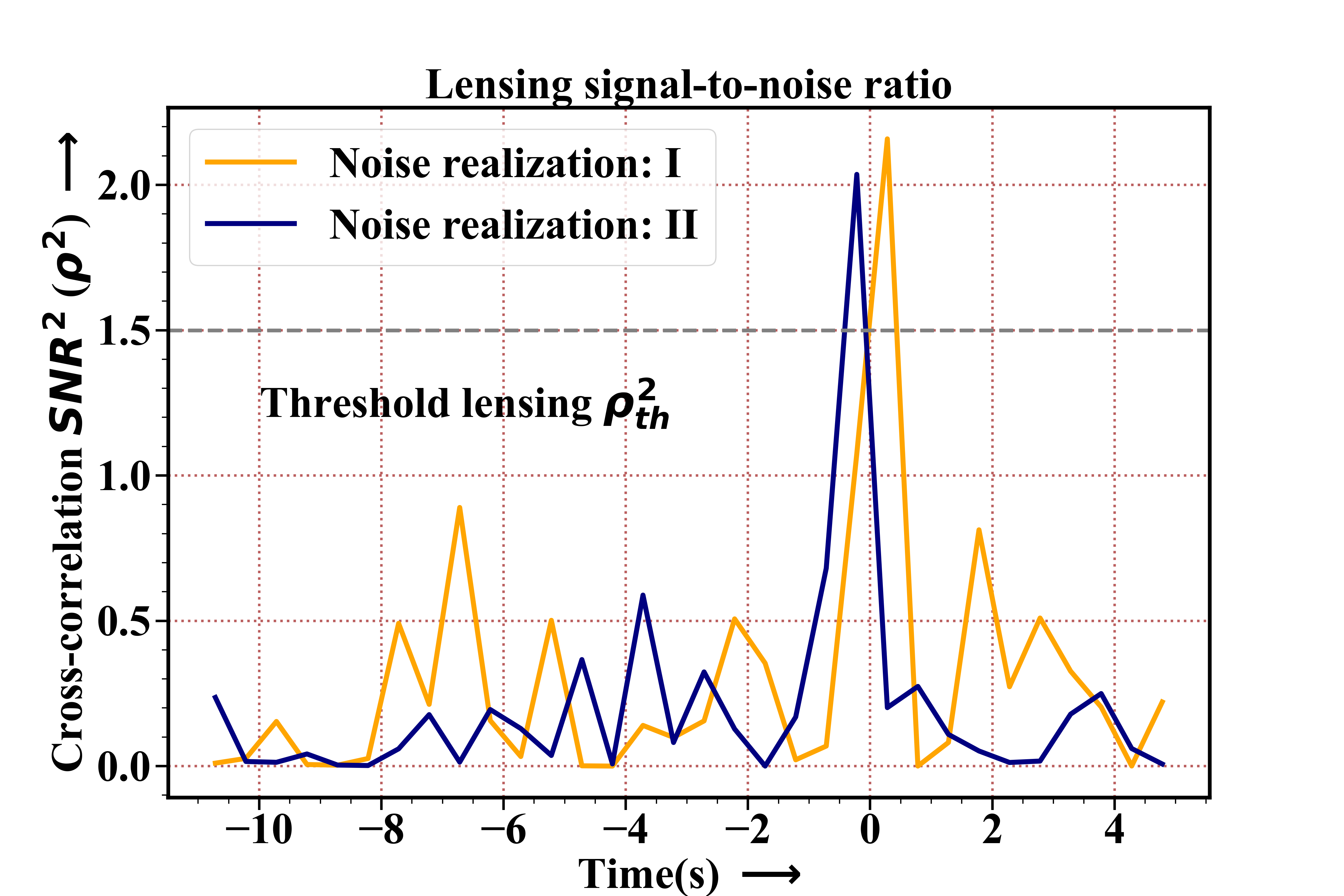}
    \caption{The figure shows the lensing SNR for two different noise realizations with same source properties as before. We can see that the threshold is lowered to $\rho^2 _{th} > 1.5 $. Although the peak corresponding to the signal (near $t=0s$) is much taller than the noise cross-correlations peaks, the overall peak height has diminished drastically compared to the cases which we showed earlier. The cut-off is chosen according to the noise fluctuations as well the SNR timescale $\tau_{snr}$, here $\tau_{snr} = 0.5s$. Thus we have no hard-bound prescription for the cut-off lensing SNR. The choice of an appropriate threshold lensing SNR ($\rho_{th}$) plays an essential role in the determination of the event significance as a lensing candidate.}
    \label{fig:appai}
\end{figure}

\begin{figure}
    \centering
    \includegraphics[width=0.45\textwidth]{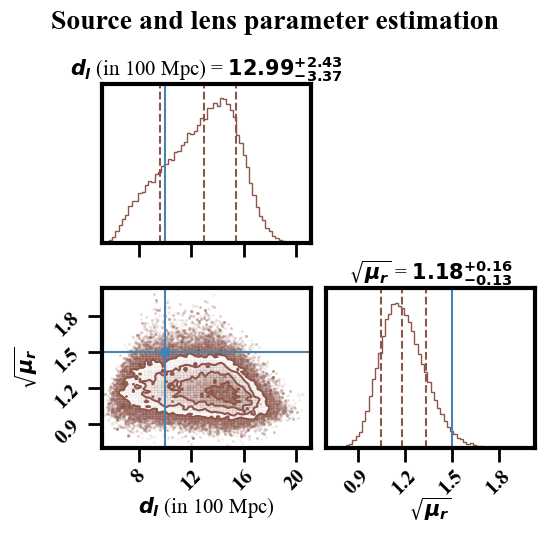}
    \caption{The figure shows the source and lens characterization for a different noise realization with same source properties as before. Although the inference of $\sqrt{\mu_r}$ is quite bad, the inference of $d_l$ is acceptable. The inference of both the parameters becomes better (with posteriors becoming narrower close to the injection value) as the strength of the signal i.e. natch-filtering SNR is increased.}
    \label{fig:appaii}
\end{figure}

As can be seen in fig. \ref{fig:dist_infer} and fig. \ref{fig:mag_infer}, the quality of the inference of the parameters depends on the signal strength. When the lensed signals have moderate match-filtered SNR, the inference standard deviation is small. However, with weak signals, parameters are estimated with large error bars. 
Not only do these kinds of estimations depend on the source parameters, but also strongly depend on the noise realizations when the signal is not that strong itself. As we tried calculating lensing SNR or running our parameter estimation on low match-filtered SNR, our results were sometimes affected badly by the noise. We have shown the variation in the cross-correlation SNR and parameter estimation for different noise realizations in fig. \ref{fig:appai} and fig. \ref{fig:appaii}.

For any peak having $\rho$ > $\rho_{th}$ in fig \ref{fig:appai} can be checked to contain any lensed signal by using the Bayesian parameter estimation technique on the data pieces as discussed in section 6. Parameter estimation does not provide a bound on the parameter if there is no signal present. This shows the importance of setting a threshold of the lensing at such a height that all peaks above the threshold are being tested thoroughly before rejection. However, we emphasize that there does not exist a universal choice of the threshold, and it can only be set by observing typical noise cross-correlation SNRs.

In fig. \ref{fig:appaii}, we can see that although the underlying noise power spectral density (PSD) is the same, however, each noise realization impacts the inference when the signal strength is weak. The effect of noise realizations on the parameter estimation goes away as we observe events with high SNR (e.g. individual detector SNR of around 10). 


\section{max SNR vs sum SNR: In the use to extract parameter information from the signal}\label{app:choicesnr}

\begin{figure}
    \centering
    \includegraphics[width=0.48\textwidth]{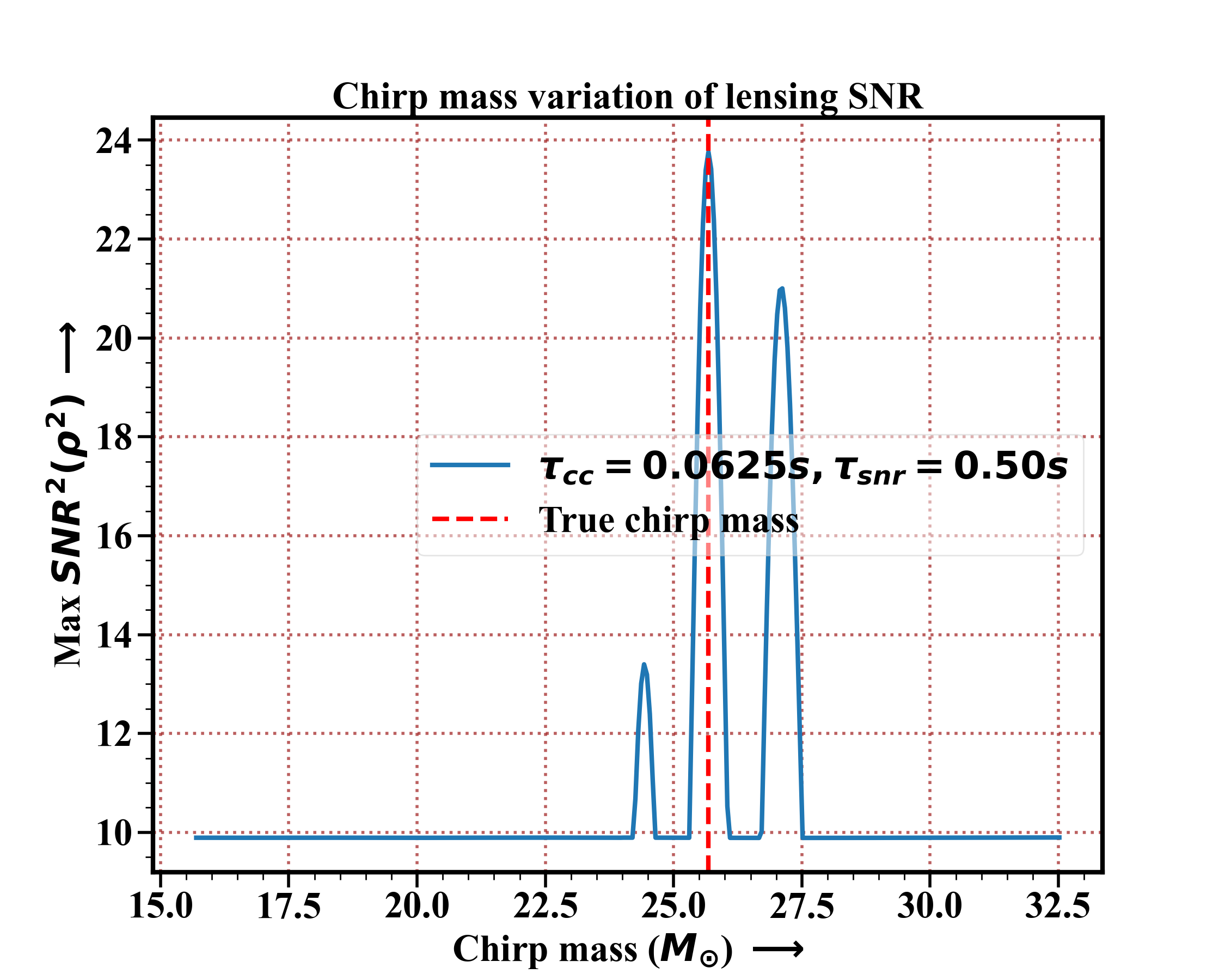}
    \caption{The figure shows the variation of max lensing SNR with chirp mass. The recovery of the injection of chirp mass is successful. The figure shows that max lensing SNR can be a proxy to summed lensing SNR to extract the source parameters accurately.}
    \label{fig:appci}
\end{figure}

\begin{figure}
    \centering
    \includegraphics[width=0.48\textwidth]{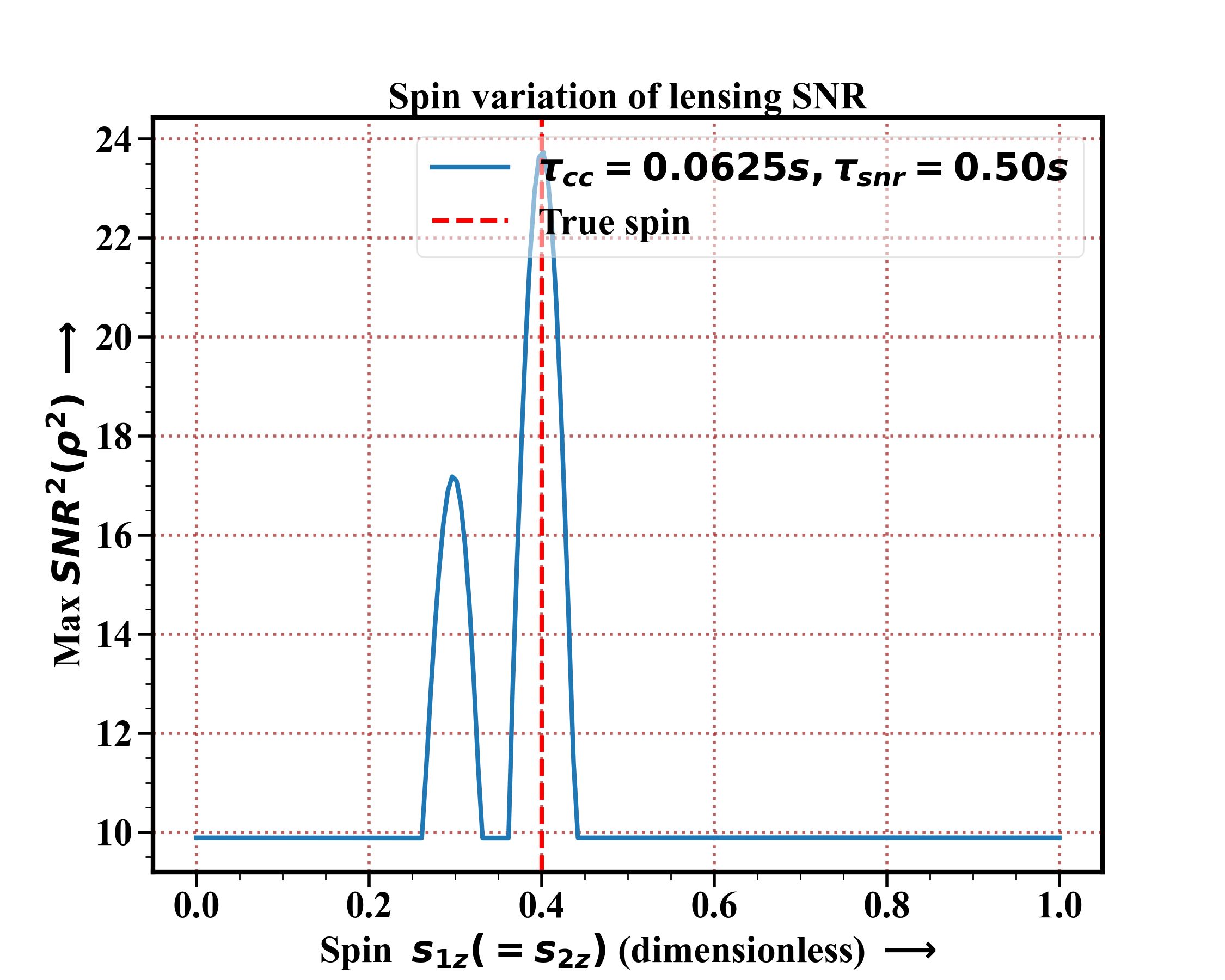}
    \caption{The figure shows the variation of max lensing SNR with component black hole spin component towards orbital plane perpendicular direction. We can correctly infer the true spins of the binary black hole from this technique by using max lensing SNR instead of summed SNR. Thus max lensing SNR can act as a replacement, if not a better one, to the summed SNR technique.}
    \label{fig:appcii}
\end{figure}

In fig. \ref{fig18}, we have shown the inference of the injected source parameter from the signal cross-correlation. We did this by observing the total lensing SNR of the cross-correlation signal and one parameter in data-II was varied keeping other parameters fixed at the values same as that in data-I. However, that is not the only way to find the value of the parameter. Instead of plotting total lensing SNR, we can plot max lensing SNR vs data-II parameter and observe where a peak is obtained. 

Fig. \ref{fig:appci} shows that the max lensing SNR can also be used to find the injected chirp mass (given an apriori knowledge of other parameters). Similarly, \ref{fig:appcii} shows the variation of the max lensing SNR in black hole component spins, which shows that it is quite possible to recover the injected spin from the max lensing SNR peak(again, given a apriori knowledge of other parameters). Thus, from the observations we can conclude that max lensing SNR can act aptly as a replacement for total lensing SNR.

For long signals, however, total (or summed) lensing SNR will play a significant role in searching for lensed event candidates because at many instances there will be a significant lensing SNR, and considering a single max lensing SNR will cause loss of information of the lensing SNRs at all other instants.

\section{Choice of Prior in parameter estimation}\label{app:g}
\begin{figure}
    \centering
    \includegraphics[width=0.48\textwidth]{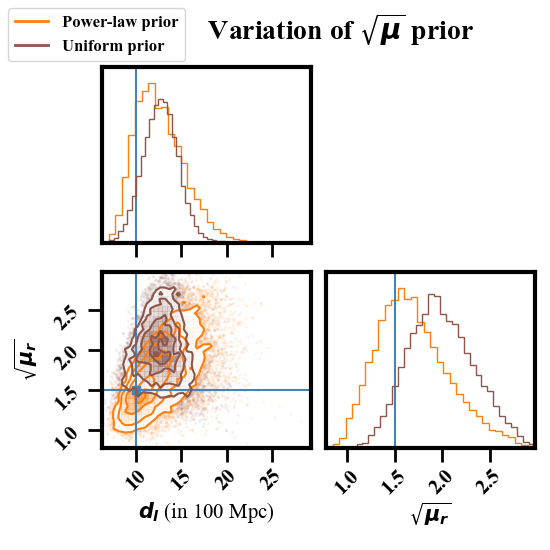}
    \caption{The figure shows the comparison of the inferences with the choice of power law vs uniform prior in magnification factor. The two inferences are very similar with the power law prior prioritizing lower $d_l$ and $\sqrt{\mu_r}$ values as compared to the uniform prior.}
    \label{fig:compare}
\end{figure}

We had chosen a flat prior for magnifications $\sqrt{\mu_i}'s$ in the parameter estimation for strong lensing. This often resulted in a railing effect, where the distributions of the magnification and luminosity distance tend to orient towards the higher end of the specified prior range. Because the choice of a uniform prior favours all values of magnification within the range equally, it occurs. Since the probability of lensed events with a magnification $\mu$ goes down with increasing magnification as $dP(\mu)/d\mu \propto 1/\mu^3$ \citep{1991ApJ...374...83R}. So, we also choose this as a physically-motivated prior on the magnification factor.  $\sqrt{\mu_i}'s$ were within the range $\sqrt{\mu_i} \in [1, 3]$, which implies a prior on $\mu_i \in [1, 9]$. 
 The luminosity distance prior is uniform and kept the same as before: $ d_l \in [ \rm{min}(d_{l1}, d_{l2}), 5\, \rm{Gpc}]$. The results with the choice of different priors in the parameter estimation are compared in the fig. \ref{fig:compare}. The figure shows that even without the choice of the physics-driven prior, the source and lens parameters can be well constrained. Although there are little variations in the observed posterior distributions, prior choice does not change its overall shape and constraints on the parameters.

\section{Choice of parameters for Cosmology model and BBH population model}\label{app:h}

To perform the false alarm rate calculation for strong lensing, we have used GWSIM \citep{Karathanasis_2023} to generate the events detectable by observatories H1, L1 and V1 within 5 years of observation with O4 noise characteristics. In the table \ref{tab:bbh_params}, we have mentioned our choice for different parameters in GWSIM the generating BBH events. We have used the merger-rate equation based on the empirical Madau-Dickinson star-formation rate given by, 
\begin{eqnarray}
    R(z)=R_0(1+z)^{m_{\alpha}} \frac{1+\left(1+z_p\right)^{-(m_{\alpha}+m_{\beta})}}{1+\left(\frac{1+z}{1+z_p}\right)^{(m_{\alpha}+m_{\beta})}}.
\end{eqnarray}

For the BBH mass-distribution model, we choose primary masses from a power-law plus Gaussian distribution given by, 
$$
\begin{array}{l} p\left(m_1 \mid M_{\min }, M_{\max }, \alpha, \mu, \sigma, \lambda, \delta \right) \propto S\left(m_1 \mid M_{\min }, \delta \right) \times \vspace{0.3cm}\\ \left\{\begin{array}{ll}\left(1-\lambda\right) P\left(m_1 \mid M_{\max }, M_{\min }, \alpha\right) \\ +\lambda G\left(m_1 \mid \mu, \sigma\right) & {\rm for \hspace{0.2cm}} M_{\min }<m_1<M_{\max }, \vspace{0.4cm}\\ 0 & \text {otherwise }\end{array}\right.\end{array}
$$
Here, $P(m_1 \mid M_{\max }, M_{\min }, \alpha)$
is the truncated power law distribution and $G\left(m_1 \mid \mu, \sigma\right)$ is a Gaussian distribution with a mean $\mu$ and standard deviation $\sigma$. The distributions are given by, 

$$
P\left(m_1 \mid M_{min}, M_{max}, \alpha \right) \propto \left\{\begin{array}{ll}
m_1^{-\alpha} & M_{\min }<m_1<M_{\max } \\
0 & \text { otherwise }
\end{array}
\right.
$$
and,
$$
G\left(m_1 \mid \mu, \sigma\right)=\frac{1}{\sigma \sqrt{2 \pi}} \exp \left(-\frac{\left(m_1-\mu\right)^2}{2 \sigma^2}\right).
$$
Here, $S(m_1 \mid M_{min}, \delta)$ is a smoothing function given by,
$$
S\left(m_1 \mid M_{\min }, \delta\right)=\left\{\begin{array}{ll}
0 & m_1<M_{\min } \\
f(m_1-M_{min}, \delta) & M_{\min }<m_1<M_{\min }+\delta \\
1 & m_1>M_{min}+\delta
\end{array}
\right.
$$
The function $f(m, \delta)$ is defined as follows,
$$
f(m, \delta)= \left(1+ e^{\left(\frac{\delta}{m} + \frac{\delta}{m-\delta} \right)} \right)^{-1}
$$
The secondary mass is sampled from a power-law model given by, 
$$
P\left(m_2 \mid M_{min}, m_1, \beta \right) \propto \left\{\begin{array}{ll}
m_2^{\beta} & M_{\min }<m_2 \leq m_1 \\
0 & \text { otherwise }
\end{array}
\right.
$$
The spins of the binary black hole are chosen from the uniform distribution given by,
$$
P\left(\chi_{\text {eff }}\right) \propto \mathcal{U}[-1,1],
$$
with both the spin orientations are either aligned to the orbital angular momentum or opposite to it.

\begin{table}
    \centering
    \begin{tabular}{|p{0.25\textwidth}| p{0.25\textwidth}|}
    \hline
        \textbf{Parameter description} & \textbf{Choice of parameter} \\ \hline
        \hline
        Population model to sample mass 1, mass 2 of BBH & Power-law plus gaussian and power-law respectively \\ \hline
        Power for the mass 1 distribution law $m_1^{-\alpha}$, $(\alpha)$ & 3.4 \\ \hline
        Power for the mass 2 distribution, law $m_2^{\beta}$, $(\beta)$ & 0.8 \\ \hline
        Maximum mass for the mass 1 distribution, ($\rm M_{max}$ in solar mass) & 100 \\ \hline
        Minimum mass for the mass 1 distribution, ($\rm M_{min}$ in solar mass) & 5 \\ \hline
        Mean of the Gaussian peak for the mass 1 distribution, ($\mu$) & 35 \\ \hline
        Standard deviation of the Gaussian peak for the mass 1 distribution, ($\sigma$) & 3.88 \\ \hline
        Weight of the Gaussian in the total distribution, ($\lambda$) & 0.04 \\ \hline
        Smoothing parameter for the mass distribution, $(\delta)$ & 4.8 \\ \hline
        Spin alignment model & Both spin-tilt angles are either 0 or $\pi$ \\ \hline
        Spin distribution model & Uniform, $P\left(\chi_{\text {eff }}\right) \propto \mathcal{U}[-1,1],$ \\ \hline
        Redshift evolution model for the merger rate & Madau-Dickinson \\ \hline
        Merger rate evolution parameter I $(m_{\alpha})$ for Madau model & 2.7 \\ \hline
        Merger rate evolution parameter II $(m_{\beta})$ for Madau model & 2.9 \\ \hline
        Peak merger rate redshift $(z_p)$ for Madau model & 1.9 \\ \hline
        Merger rate in $\rm{Gpc}^{-3}.\rm{year}^{-1}$ at $z=0$ $(R_0)$ & 20 \\ \hline
    \end{tabular}
    \caption{The table consists of our choice for  the models and parameters used in the BBH merger-rate, source-masses and spins.}
    \label{tab:bbh_params}
\end{table}

\bsp	
\label{lastpage}
\end{document}